\documentclass{pasa}
 \usepackage[
	pdftex,
	pdfpagemode={UseOutlines},
	bookmarks,
	bookmarksopen,
	colorlinks,
	linkcolor={blue},
	citecolor={green},
	urlcolor={red}
]{hyperref}
 \usepackage[english]{babel}
 \usepackage{graphicx}
 \usepackage[authoryear]{natbib}
 \usepackage{amssymb}
 \usepackage{amsmath}
 \usepackage{multirow}
 \usepackage{dcolumn}
 \bibpunct{(}{)}{;}{a}{}{,}
 \graphicspath{{ART/}}
 \DeclareGraphicsExtensions{.pdf,.png,.jpg}
 \newcolumntype{d}[1]{D{.}{.}{#1}}
 \newcolumntype{.}{D{.}{.}{-1}}
 \newcolumntype{,}{D{,}{,}{2}}
 \date{}
\def\smartspace#1{{\protect\aftergroup\smartspaceit#1}}
\def\smartspaceit{\futurelet\spta\sptest}
\def\sptest{\ifcat\noexpand\spta,\else\ \fi}

\newcommand{\nc}{\newcommand}

\newcommand{\esm}{\ensuremath}
\newcommand{\mcl}{\multicolumn}
\newcommand{\sspc}{\smartspace}
\nc{\etal}{\sspc{et~al.}}
\nc{\eg}{\sspc{e.g.}}
\nc{\etc}{\sspc{etc.}}
\nc{\cf}{\sspc{cf.}}
\nc{\ie}{\sspc{i.e.}}
\nc{\cfig}[1]{\centerline{\psfig{#1}}}
\nc{\Ncol}[1]{\mcl{#1}{c}{~}}
\nc{\mcn}[1]{\mcl{#1}{l}{~}}
\nc{\mcN}[2]{\mcl{#1}{c}{#2}}
\nc{\mcc}[1]{\mcl{1}{c}{#1}}
 \nc{\AAA}{\sspc{\esm{\lambda\lambda}}}
 \nc{\amm}{\sspc{\,\AA\,mm$^{-1}$}}
 \nc{\kms}{\sspc{\,\esm{\text{km s}^{-1}}}}
 \nc{\msun}{\sspc{\,\esm{\text{M}_{\odot}}}}
 \nc{\rsun}{\sspc{\,\esm{\text{R}_{\odot}}}}
 \nc{\lsun}{\sspc{\,\esm{\text{L}_{\odot}}}}
 \nc{\yr}{\sspc{\,\esm{\text{yr}}}}
 \nc{\kpc}{\sspc{\,\esm{\text{kpc}}}}
 \nc{\halpha}{\sspc{\,\mbox{H$\alpha$}}}
 \nc{\hbeta}{\sspc{\,\mbox{H$\beta$}}}
 \nc{\hii}{\rm H\,{\sc ii} }
 \nc{\ion}[2]{\sspc{#1\,{\scshape#2}}}
 \nc{\Ion}[2]{\sspc{#1\,#2}}
 \nc{\Line}[3]{\sspc{[#1]$\lambda$#2/$\lambda$#3}}
 \nc{\dgr}{\esm{^\circ}}
 \nc{\arcmin}{\sspc{\,\esm{\mkern-4mu^\prime}}}
 \nc{\arcsec}{\sspc{\,\esm{\mkern-4mu^{\prime\prime}}}}
 \nc{\Mag}[2]{#1\fm#2}
 \nc{\dex}[1]{\esm{10^{#1}}}
 \nc{\tdex}[1]{\esm{\times10^{#1}}}
 \nc{\todo}[1]{\sspc{(\textbf{TODO:} #1)}}
 \nc{\obj}[1]{\sspc{#1}}
 \nc{\rotsed}{\sspc{ROTSE--IIId}}
 \nc{\rotse}{\sspc{ROTSE--III}}
\title
[%
  Fluxes of Northern PNe%
]{%
  Emission-Line Fluxes of Northern Planetary Nebulae%
}
\author
[%
  Aksaker et.al.%
]{%
  N.\     Aksaker$^{1}$\thanks{Correspondence address: naksaker@cu.edu.tr},
  S.\,K.\ Yerli$^{2}$,
  \"U     K{\i}z{\i}lo\u{g}lu$^{2}$,
  \and
  B.\     Atalay$^{3}$ \\
\affil{$^{1}$ Vocational School of Technical Sciences,  University of \c{C}ukurova, Adana, Turkey}%
\affil{$^{2}$ Orta Do\u{g}u Teknik \"Universitesi, Department of Physics, Ankara, Turkey}%
\affil{$^{3}$ Atat\"urk University, Faculty of Science, Department of Physics, Erzurum, Turkey}%
}
\jid{PASA}
\doi{10.1017/pas.\the\year.xxx}
\jyear{\the\year}
\begin{document}
\begin{abstract}
We present long slit spectrophotometric emission line fluxes of bright and extended ($<$5 arcsec in diameter) Planetary Nebulae (PNe) selected from \citealp{1992secg.book.....A} catalog with suitable equitorial coordinates for Northern hemisphere. In total, 17 PNe have been choosen and observed in 2008--2010. To measure absolute fluxes, broad slit sizes, ranging from 3.5\arcsec to 7.5\arcsec were used and thus equivalent widths of all observable emission line fluxes were also calculated. Among 17 PNe's observed, line flux measurements of 12 of them were made for the first time.
This work also aims to extend the sky coverage of emission line flux standards in Northern hemisphere (\citealp{1997ApJS..108..515D} - 52 PNe in Southern hemisphere; \citealp{2005A&A...436..967W} - 6 PNe in Northern hemisphere).
Electron temperatures and densities, and chemical abundances of these PNe were also calculated in this work.
These data is expected to lead the photometric or spectrometric further work for absolute emission line flux measurements needed for \hii regions, supernova remnants etc.
\end{abstract}


\maketitle
\section{Introduction}
\label{intro}

Studying the stellar evolution in terms of chromospheric activities and kinematics of stars themselves within the Galaxy could also impact on what we know about planetary nebulae (PNe) \ie when the star's interiors are completely revealed with enriched matter content which will be fed to interstellar medium (ISM) with stellar wind, and therefore enriching the ISM.
The key point in these studies lie in the spectroscopic observations of PNe when accurate emission line fluxes are measured (\citealp{2006IAUS..234..455M}).
Moreover, with the help of this knowledge we can understand how these regions are related to the general structure of the ISM.

In the past there were several works that they have relied on standard star observations using broad band photometry or spectrophotometry, or even using photon counting detectors.
However, these observations help little in improving the emission line flux standards of PNe, HII regions, cataclysmic variables, supernova remnants, etc. (\citealp{2005A&A...436..967W}).
For photon counting coupled with interference filters: \citealp{1955ApJ...122..240L, 1963ApJ...138.1018O}; they had several large error sources such as unknown central wavelength of transmission curves, temperature variations, filter age and misalignment of various optical components.
For broad band photometry (\citealp{1954ApJ...120..196J, 1992AJ....104..340L}) and spectrophotometry: (\citealp{1983MNRAS.204..347S, 1988ApJ...328..315M, 1990AJ.....99.1621O}); even though there are many type of standard stars available, the flux standards of emission lines are still not adequate in number. In addition to this, several emission and absorption lines coincide in continuum standards (e.g. Balmer series) and when narrow band filters are used, exact contribution of the line fluxes would be difficult.
Contrary to abovementioned observation methods, using low read-out noise and high quantum efficiency CCDs, spectrophotometry of emission lines of PNe's could very accurately be measured even for faint sources.

Therefore, high precision photometric studies of the emission line fluxes of ionized nebulae are needed.

Lately, \citealp{1997ApJS..108..515D} (hereafter D97) made slitless spectrophotometric observation of southern compact PNe chosen from the catalog of \citealp{1992secg.book.....A} (hereafter A92). They gave \halpha, \hbeta, \ion{O}{III} and laboratory wavelengths of \ion{N}{II} and \ion{S}{II} emission lines for 39 PNe.
For northern hemisphere sources \citealp{2005A&A...436..967W} (hereafter W05) presented all emission lines of only 6 PNe using similar techniques. Although this was off to a good start, it is not sufficient for northern hemisphere.

By increasing the northern hemisphere sources to 17 with this work, a complete set of emission line standards will be builded up for both hemisphere.
\section{Observations}
\label{sec:obs}

The catalog of A92 includes 1142 PNe. Therefore, it is used as the main source of our work. The PNe to be observed were selected according to the following criteria:
\begin{itemize}
\item To select only northern hemisphere PNe, first declination is limited to $\delta > -35 \dgr$ which reduces the total number to 838.
\item Then, RA in between 7$^h$ and 12$^h$ are excluded from the selection which reduces the total number to 805 PNe.
\item From this subset, angular size of PNe smaller than 5 arcsecond are selected (\ie PNe fits to the available slit size). This reduced the total number to 276 PNe.
\item Then, less studied 12 PNe were selected.
\item 5 PNe which were studied by W05 were added to this subset increasing the total number to 17.
\end{itemize}
Journal of observations for the selected PNe are given in Table \ref{tab:1}.
\begin{table*}
 \caption{Journal of Observations for the target list. Description of the columns are as follow. The most common object names (Column 1), Source coordinates (Column 2 and 3), R magnitude of the source given in SIMBAD  (Column 4), Source diameter in arcsec (Column 5) given in A92 catalog, Total exposure for blue and red grism (Column 6) and Observation date (Column 7)}
\label{tab:1}
\begin{small}
\begin{center}
\begin{tabular}{@{}l@{~}l@{~}c@{~}c@{~}c@{~}c@{~}c@{~}}
\hline
Object Name
& $\alpha_{2000}$
& $\delta_{2000}$
& R Mag.
& Diameter
& Exposure
& Obs. Date
\\
& (h m s)
& (d m s)
&
& (\arcsec)
& (s)
\\
\hline
Hu 1-1&00 28 15.44&+55 57 54.48&-&5&3000&10 Jul. 2010\\
BoBn 1&00 37 16.03&-13 42 58.60&14.6&3&2400&9 Nov. 2008\\
M 1-4&03 41 43.37&+52 17 00.54&11.9&4&1800&11 Jul. 2010\\
M 1-5&05 46 50.01&+24 22 02.80&12&2&2400&9 Nov. 2008\\
K 3-70&05 58 45.34&+25 18 43.80&13.9&1.8&3600&9 Nov. 2008\\
M 1-17&07 40 22.20&-11 32 29.92&12.4&3&1200&10 Nov. 2008\\
SaSt 2-3&07 48 03.67&-14 07 40.40&13.1&0&1200&10 Nov. 2008\\
H 4-1&12 59 27.77&+27 38 10.50&14&2.7&6600&11 Jul. 2010\\
DdDm 1&16 40 18.15&+38 42 20.00&12.9&0.6&2400&9 Nov. 2008\\
Na 1&17 12 51.89&-03 15 59.69&10.9&5&4200&10 Jul. 2010\\
Vy 1-2&17 54 22.98&+27 59 58.10&12.4&4.6&4200&10 Jul. 2010\\
M 3-27&18 27 48.26&+14 29 06.10&11.9&1&2100&11 Jul. 2010\\
NGC 6833&19 49 46.58&+48 57 40.20&11&2&1800&10 Nov. 2008\\
IC 4997&20 20 08.74&+16 43 53.71&-&1.6&2100&11 Jul. 2010\\
IC 5117&21 32 30.97&+44 35 47.50&-&1.2&780&10 Nov. 2008\\
Me 2-2&22 31 43.68&+47 48 03.91&10.8&$<$5&180&11 Jul. 2010\\
Vy 2-3&23 22 57.89&+46 53 57.79&12.1&4.2&3600&10 Jul. 2010\\
 \hline
\end{tabular}
\end{center}
\end{small}
\end{table*}

\begin{figure*}
\centerline{\includegraphics[width=0.7\textwidth]{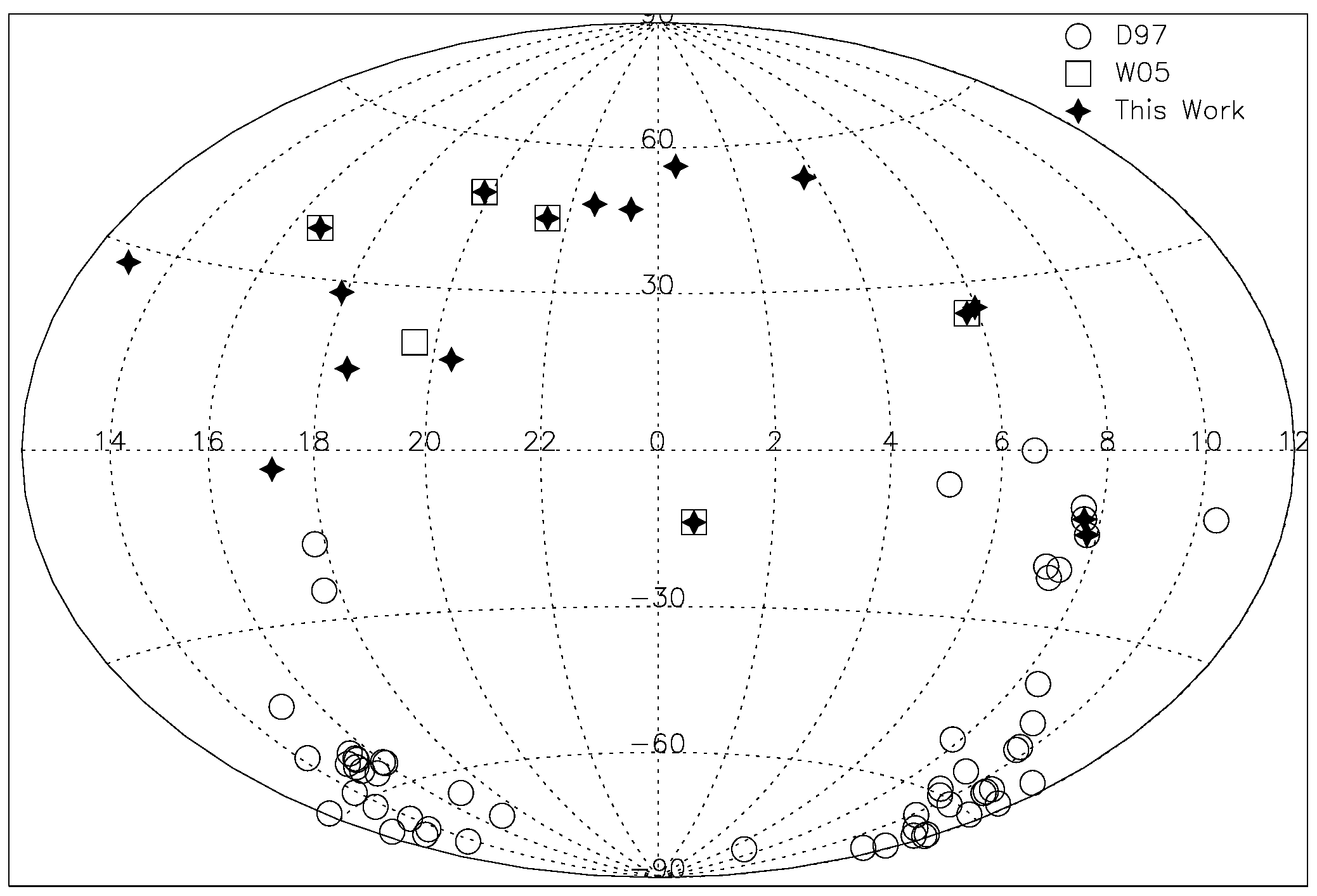}}
\caption{The sky distribution of our target list (stars) in equatorial coordinates. The source of W05 and D97 are represented by squares and circles, respectively.}
\label{f:coverage}
\end{figure*}

Sky distribution of target list given in Table \ref{tab:1} is shown in Figure \ref{f:coverage}. It contains sources of both D97 and W05. It can easily be seen from the figure that our aim of filling sky coverage gaps in northern hemisphere is achieved.

The observations were performed with the TFOSC\footnote{http://tug.tubitak.gov.tr/tr/teleskoplar/tfosc} (T\"UB\.ITAK Faint Object Spectrometer and Camera) coupled with the 150 cm Russian-Turkish Telescope (RTT150). The camera is equipped with a 2048x2048 (15 $\mu$m) pixel Fairchild 447BI CCD.

The mean seeing level through out the observing nights were 1.9\arcsec, ranging from 0.6\arcsec to 2.6\arcsec. The seeing measurements were calculated from unfiltered frames by averaging over FWHM's of field stars which were taken just before the spectrum observations. The seeing characteristics of the site were well determined by \cite{2004A&A...422.1129O}.

In the spectroscopic work, mainly G7 (Grism 7: 3850-6850 \AA; visible), G8 (Grism 8: 5800-8300 \AA; red) and G14 (Grism 14: 3275-6100 \AA; blue) were used. The average dispersions were 1.5, 1.1, and 1.4 \AA/pixel for G7, G8, and G14, respectively. Considering the telescope optics and seeing conditions, an optimum Signal-to-Noise Ratio (SNR) was reached with the following exposure times: 60 s for {\it Me 2-2} (a typical bright PNe) and 3600 s for {\it H 4-1} (a typical faint PNe). Note that the exposure times were determined according to source's R magnitude given in Table \ref{tab:1}.

Each PNe was observed with a slit size which was varied according to the PNe's angular size and seeing: 3.5\arcsec for point-like sources and 7.3\arcsec for wider sources. The slit position and therefore dispersion axis is aligned to the parallactic angle to avoid loss of light due to the atmospheric refraction. Contrary to work of D97 in southern hemisphere, PNe coordinates in northern hemisphere were quite accurate: Source-to-source pointing accuracy was kept $<$ 5\arcsec and there were no loss of frames due to the autoguider failures.

\section{Data Analysis}
\label{sec:data}

Headers of TFOSC images were processed using IDL%
\footnote{http://www.exelisvis.com/language/en-US/ProductsServices/IDL.aspx}
and then they were reduced using standard IRAF%
\footnote{IRAF  is distributed by the National Optical Astronomy Observatories, which  are operated by the Association of Universities for Research in  Astronomy, Inc., under cooperative agreement with the National Science  Foundation.}
tasks \eg \textit{onedspec} and \textit{image}. The instrumental profile of the sensor was subtracted using standard reduction packages of IRAF. The images were then converted into 1D spectra by using the \textit{apall} task. As for the last stage wavelength and flux calibrations tasks were applied to the spectra.

To achieve a sensible accuracy in the wavelength calibration, different arc lamps were used for different grisms: He+Ne for Grism 7 (hereafter G7); Ne for Grisim 8 (hereafer G8); He for Grism 14 (hereafter G14). Arc lamp atlas of ALFOSC (another spectrograph in FOSC series)\footnote{http://www.not.iac.es/instruments/alfosc/lamps/} were used to identify the lines in the spectra. The resultant accuracy of wavelength calibration was around $\pm$1 {\AA}. The reference wavelength used in flux measurements are given in Tables \ref{tab:3}-\ref{tab:19}.

Fluxes of spectrophotometric standards listed in \cite{1983MNRAS.204..347S}, \cite{1984MNRAS.206..241B}, \cite{1988ApJ...328..315M} and \cite{1990AJ.....99.1621O} were used in flux calibration of all PNe. The standard stars have to be identified both in one of the abovementioned catalogs and in IRAF's repository of standard stars; then the closest one to the target PNe was chosen to eliminate the effect of airmass. Finally, flux calibration tasks of IRAF (\textit{standards}, \textit{sensitivity}, and \textit{calibrate}) applied to PNe images.

\begin{figure*}
\begin{center}
\begin{tabular}{@{}c@{~}}
\includegraphics[width=0.7\textwidth]{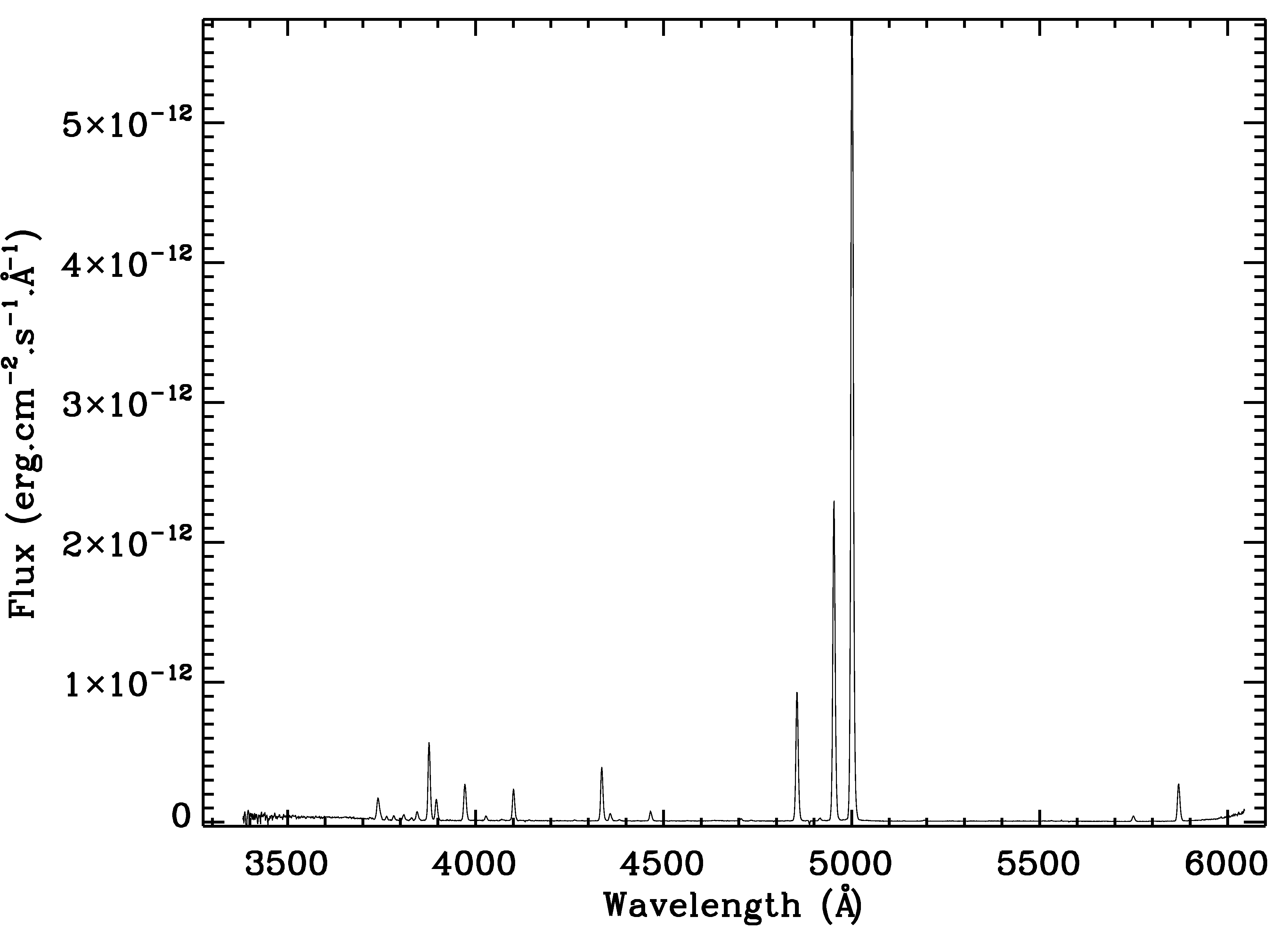}\\
\includegraphics[width=0.7\textwidth]{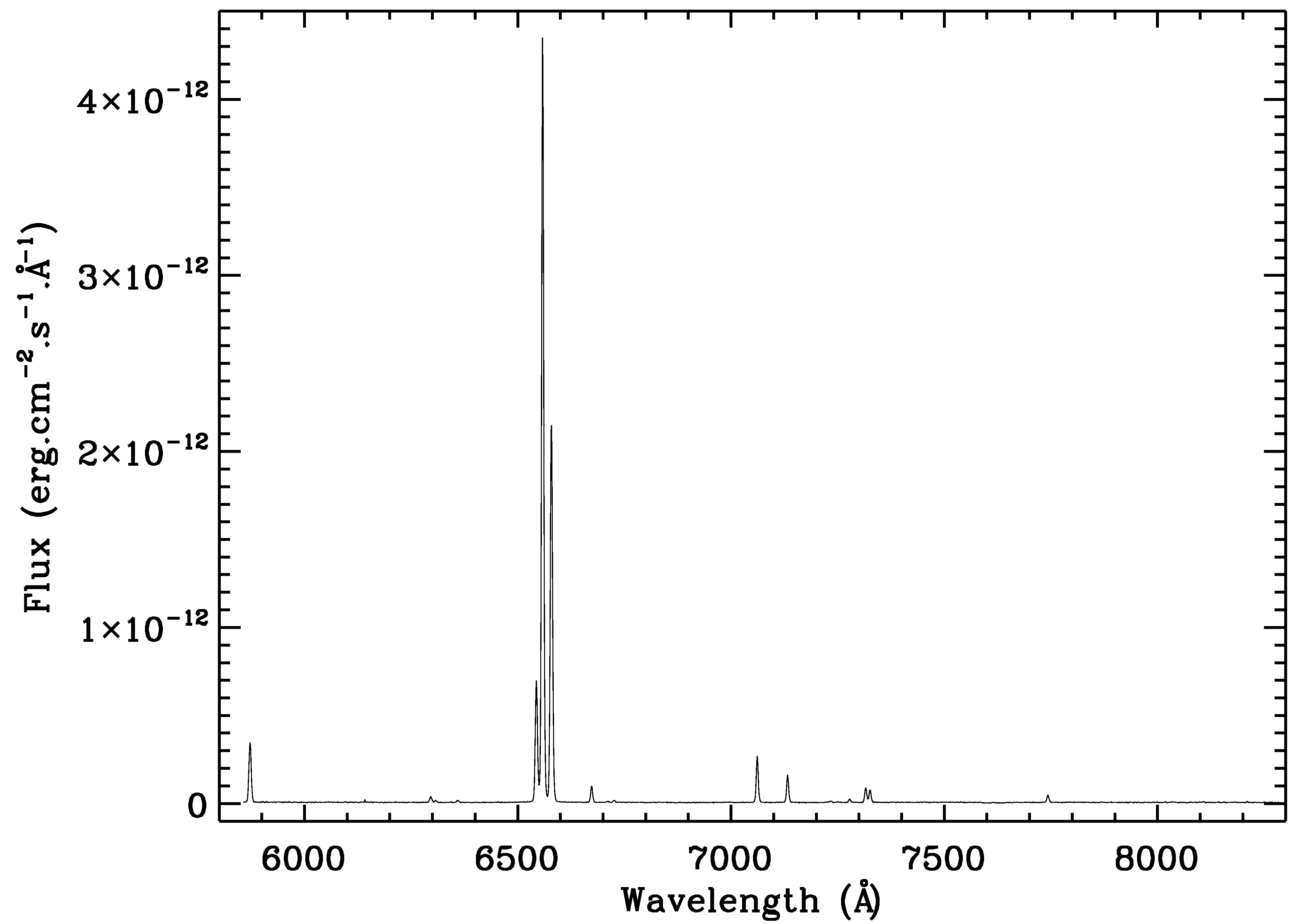}
\end{tabular}
\caption{The blue (above panel) and red (below panel) spectra of Me 2-2.}
\label{fig:F2}
\end{center}
\end{figure*}

A typical reduced pair (red and blue) of spectra are shown in Figure \ref{fig:F2}. The bright lines in red spectrum are \ion{N}{II} $\lambda$6548.05, \halpha $\lambda$6582.85 and \ion{N}{II} $\lambda$6583.45, and in blue spectrum \hbeta $\lambda$4861.29, \ion{O}{III} $\lambda$5007.57 and \ion{O}{III} $\lambda$4959.52.
\section{Results And Discussions}
\label{sec:res}

The data set of each PNe is produced by measuring line flux and their equivalent widths of selected emission lines. In measuring these values background of each emission line has to be determined and then removed by a first order polynomial fit to the line extending to its wings. Background level of each emission line on the continuum was determined around the line.

The flux values and equivalent widths were then calculated by best fitting to the line. The values and their errors are given in Tables \ref{tab:3}-\ref{tab:19} which were computed by averaging multiple measurements of Gaussian fits of the same line. The standard deviation of the fits are then taken as the error values. All the measurements were done using standard tasks of IRAF.

Our values and values of W05 data for PNe: BoBn 1, M 1-5, NGC 6833 and IC 5117 are given in Tables \ref{tab:3}-\ref{tab:6}, respectively. Missing values in the tables are due to the grisms having different wavelength coverages. For PNe DdDm 1, however, due to having no overlapping wavelength coverage it is excluded in Table \ref{tab:13}. Note that, \ion{He}{I} lines are double blended in all data sets.

The corresponding error values of the data given in Tables \ref{tab:3}-\ref{tab:19} are around $\pm$0.03 dex with a standard deviation of $\pm$0.003 dex. To get an impression of the error budget, percentages of flux-to-error ratios are summed over both for all the emission lines and for five bright emission lines, and they are listed in Table \ref{tab:2}.
\begin{table}
\caption{The error budget of the measurements of all 17 PNe: for all emission lines (first column) and five bright lines of \halpha, \hbeta, \ion{O}{III}, \ion{N}{II} and \ion{He}{I} (second column).}
 \label{tab:2}
\begin{small}
\begin{center}
\begin{tabular}{@{}l@{~}r@{~}r@{}}
\hline \hline
   \mcc{Object Name}
 & \mcc{Total error}
 & \mcc{Total error}
  \\
  \mcc{}
 & \mcc{(all lines)}
 & \mcc{(bright lines)}
  \\
\hline
Hu 1-1&0.2\%&$<$0.1\%\\
BoBn 1&0.6\%&0.3\%\\
M 1- 4&0.4\%&0.3\%\\
M 1- 5&0.5\%&0.2\%\\
K 3-70&0.2\%&0.1\%\\
M 1-17&0.1\%&$<$0.1\%\\
SaSt 2- 3&$<$0.1\%&$<$0.1\%\\
H 4- 1&0.2\%&0.1\%\\
DdDm 1&0.2\%&$<$0.1\%\\
Na 1&0.4\%&0.1\%\\
Vy 1- 2&0.2\%&$<$0.1\%\\
M 3-27&0.3\%&0.2\%\\
NGC 6833&0.3\%&$<$0.1\%\\
IC 4997&0.3\%&$<$0.1\%\\
IC 5117&0.2\%&$<$0.1\%\\
Me 2-2&0.2\%&$<$0.1\%\\
Vy 2- 3&0.1\%&$<$0.1\%\\
 \hline
 \end{tabular}
\end{center}
\end{small}
\end{table}

Number of lines detected for each object varied between 11 to 35 amounting to 375 emission lines for all objects. In addition to this in total 49 \textit{different} emission lines were identified where they were mostly caused by collisions \citep[p.260]{2000oepn.book.....K} or they were the recombination lines of \ion{H}{I}, \ion{He}{I} and \ion{He}{II}. However, a considerable amount of recombination lines of heavy elements were also detected such as \ion{O}{I}, \ion{O}{II}, \ion{O}{III}, \ion{C}{IV} and \ion{Ne}{III}.
In some of the sources (BoBn 1, K 3-70, DdDm 1 and M 1-4) emission lines were observed on an elevated continuum level which was due to the recombination of electrons.

Among 49 emission lines, the brightest (unsaturated) and faintest ones were the \ion{O}{III} $\lambda$5007.57 with $-9.947\pm0.002$ from IC4997, and \ion{N}{II} $\lambda$6548.05 with $-15.115\pm0.080$ from M 1-5, respectively. Emission line fluxes of PNe are directly proportional with optical R magnitude of the source (see Table \ref{tab:1}) which was also noted in this work (see Tables \ref{tab:3}-\ref{tab:19}).

\begin{figure}
\begin{center}
\centerline{\includegraphics[width=0.4\textwidth]{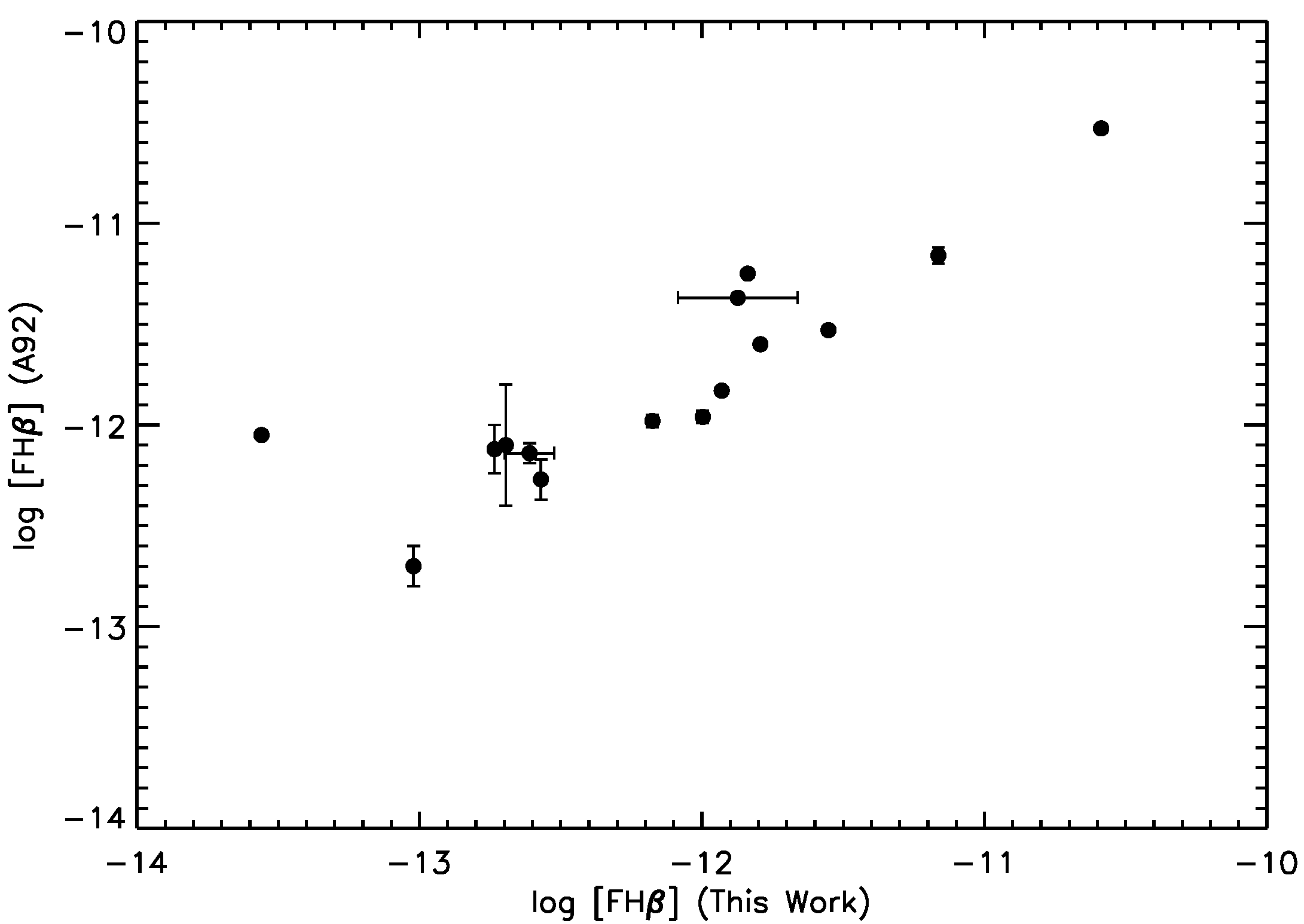}}
\centerline{\includegraphics[width=0.4\textwidth]{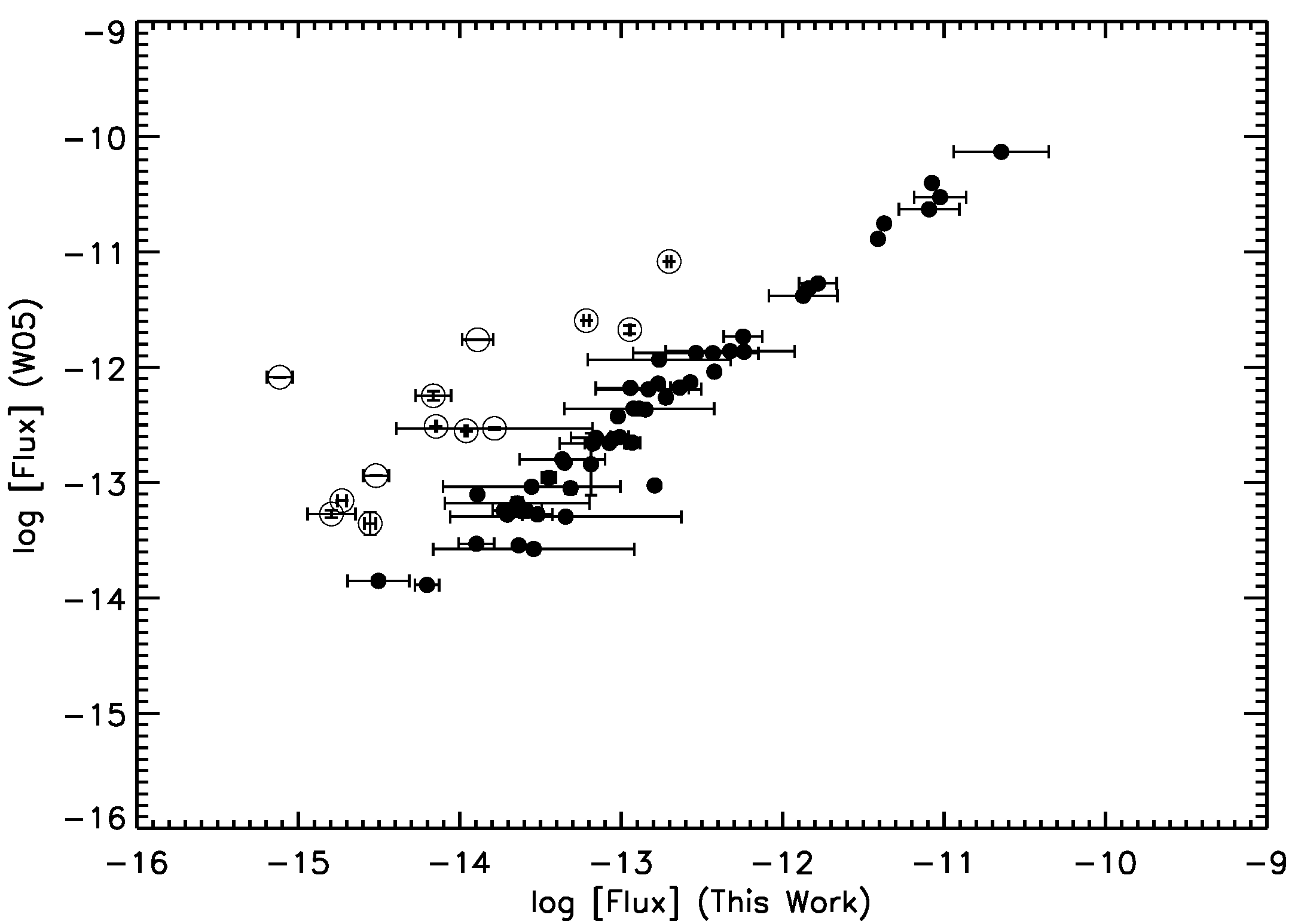}}
\centerline{\includegraphics[width=0.4\textwidth]{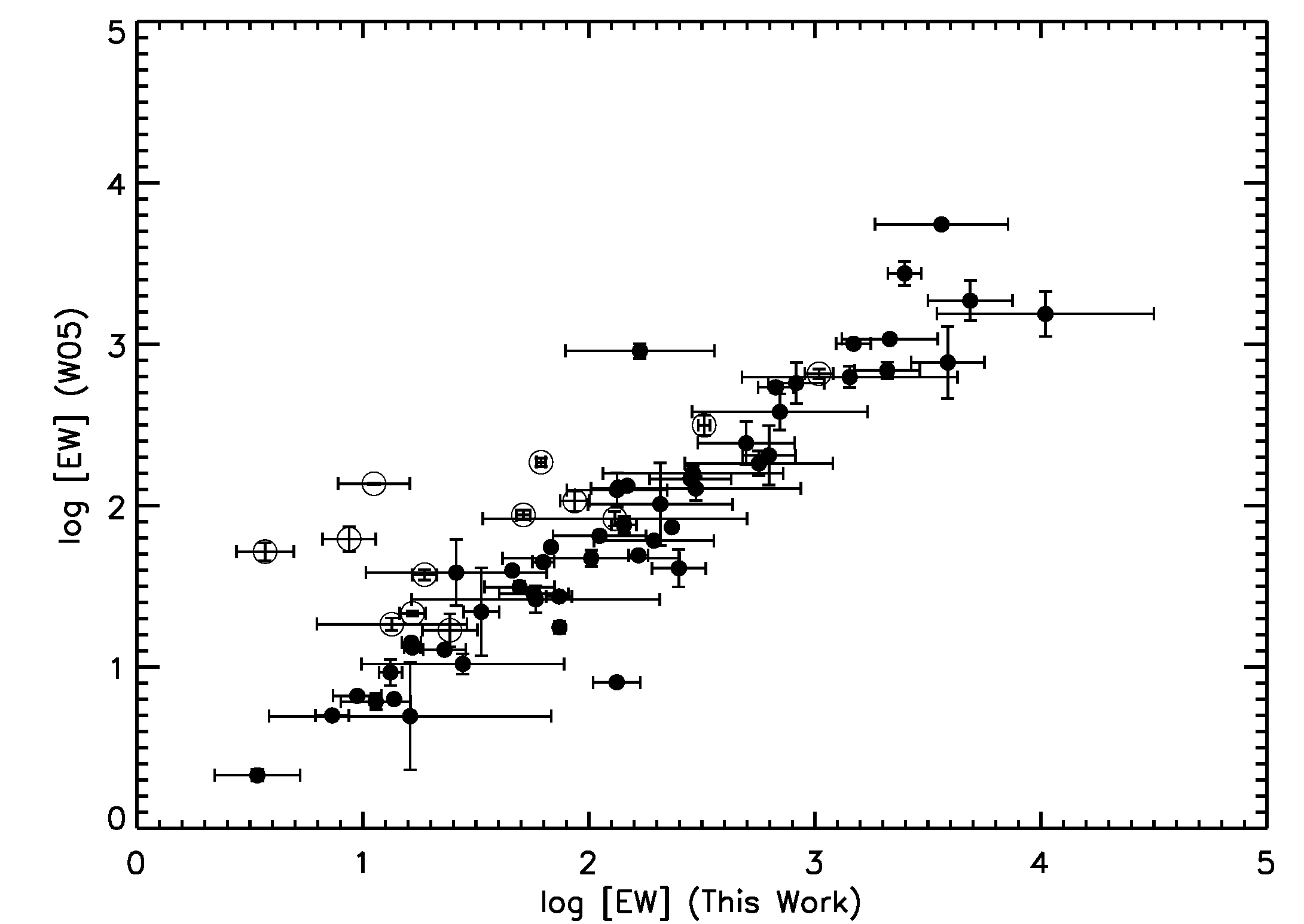}}
\caption{In the top panel a comparison of \hbeta fluxes of A92 (vertical) with this work (horizontal) is given. In the middle and bottom panels, a comparison of all emission line fluxes (in total 64 for 4 PNe) and EW of W05 (vertical) with this work (horizontal) are given, respectively. Open circles represent faint and blended \ion{He}{I} lines. All axes are in log scale and fluxes are in $ergs cm^{−2} s^{-1}$.}
\label{fig:F3}
\end{center}
\end{figure}

Observed \hbeta fluxes were compared with A92 and plotted in top panel of Figure \ref{fig:F3} (17 in total). As can be deduced from the plot, an obvious linear relation exists for both bright and faint flux arms with a R$^{2}$ value 0.89 on the faint arm.

A similar comparison was done for W05 including all emission lines (64 in total). However, only 4 PNe (BoBn 1, IC 5117, M 1-5, and NGC 6833) were used in the comparison (see Tables \ref{tab:3}-\ref{tab:6}). The comparisons for flux and EW values are shown in middle and bottom panels of Figure \ref{fig:F3}, respectively. Corresponding R$^{2}$ values of flux and EW are 0.96 and 0.88, respectively.
The deviations seen in the regression are mainly due to not resolving the faint lines.

In comparing our values with W05, there are some discrepancies which have to be mentioned in detail.
The error in flux measurement of each emission line (Table 2) amounts to less than 1\% of the line strength. On top this, low SNR of the spectra on relatively bad seeing conditions produced slit losses which accumulated to no more than 10\% of the signal, therefore causing a relatively low flux value. However, in the absence of these negative effects a high consistency is achieved in the comparison which makes the results on the whole reliable.

Physical and chemical properties for all PNe having suitable data sets were also calculated.
The analysis of the data was based on \citealp{1997A+A...322..975C}.
Since the Balmer decrement is a sign of extinction along the observed column of ISM, extinction constant ($c_\beta$) were calculated using \Ion{H}{$\alpha$}/\Ion{H}{$\beta$} ratio.
Electron temperatures ($T_{\rm e}$) and densities ($N_{\rm e}$) were calculated using IRAF's \textrm{nebular.temden} task.
Electron temperatures of $T_{\rm e}$([O III]) and $T_{\rm e}$([N II]) were calculated according to \cite{2006agna.book.....O}.
Using ratio of \Line{\ion{S}{II}}{6716}{6731} and assuming an electron temperature of T=$10^4$ K, electron densities were calculated.
All above calculations are given in Table \ref{t:chem}.
\begin{table}
 \caption{Logarithmic extinction constant $c_\beta$, and electron temperatures and densities.}
\label{t:chem}
\begin{small}
\begin{center}
\begin{tabular}{@{}l@{~}r@{~}r@{~}r@{~}r@{}}
\hline
Object
& \mcc{$C_{\beta}$}
& \mcc{$T_{\rm e}$([O III])}
& \mcc{$T_{\rm e}$([N II])}
& \mcc{$N_{\rm e}$([S II])}
\\

&
& \mcc{(K)}
& \mcc{(K)}
& \mcc{(cm$^{-3}$)}
\\
\hline
Hu 1-1   & 1.32 & 11~300 & -      & -      \\
BoBn 1   &-     & 12~300 & -      & -      \\
M 1-4    & 2.20 & -      & -      &  1~100 \\
M 1-5    & 1.24 & -      & 20~300 & -      \\
K 3-70   &-     & -      & 12~200 &  2~100 \\
M 1-17   & 1.03 & -      & 11~900 &  4~000 \\
SaSt 2-3 & 0.40 & -      & 10~000 &  2~700 \\
H 4-1    & 0.18 & 12~200 &  9~400 &    500 \\
DdDm 1   &-     & -      & -      &  2~100 \\
Na 1     & 0.94 & 10~500 & -      & -      \\
Vy 1-2   &-     & 11~700 & -      &  3~700 \\
M 3-27   & 2.68 & -      & -      & -      \\
NGC 6833 & 0.04 &  6~000 & 29~600 &$>$10~000 \\
IC 4997  & 0.30 & -      & 15~900 &$>$10~000 \\
IC 5117  & 1.22 &  9~900 & 23~700 &$>$10~000 \\
Me 2-2   & 0.37 & 10~500 &  4~500 &  1~700 \\
Vy 2-3   & 0.34 & 10~400 & -      & -      \\
 \hline
\end{tabular}
\end{center}
\end{small}
\end{table}

For each PNe, the ionic abundances relative to Hydrogen are computed from the line fluxes relative to \hbeta using the temperatures and densities values given in Table \ref{t:chem} (see \eg \citealp{1994MNRAS.271..257K,2005MNRAS.362..424W} for computation methods). These abundances are given in Table \ref{t:ion}.
\begin{table}
\caption{Ionic abundances relative to Hydrogen. All values are scaled with $10^{-6}$.}
\label{t:ion}
\begin{small}
\begin{tabular}{@{}l@{~~}c@{~~}c@{~~}c@{~~}c@{~~}c@{~~}c@{~~}c@{}}
\hline
Object
& \ion{O}{I}
& \ion{O}{III}
& \ion{N}{I}
& \ion{N}{II}
& \ion{S}{II}
& \ion{S}{III}
& \ion{Ar}{IV}
\\ \hline
M 1-5    & 1.28 & -    & -    & 5.14 & -    & -    & -   \\
M 1-17   & 43.8 & 195  & 3.08 & 6.16 & -    & 7.30 & -   \\
SaSt 2-3 & 18.1 & 1.08 & -    & 13.5 & -    & -    & -   \\
H 4-1    & 31.7 & 249  & 2.15 & 18.6 & 6.77 & -    & -   \\
NGC 6833 & 0.44 & 1770 & -    & 16.2 & 0.22 & 0.16 & -   \\
IC 4997  & -    & 33.7 & -    & 2.09 & -    & 2.05 & 0.06\\
IC 5117  & 3.06 & 29.0 & -    & 0.69 & 0.33 & 1.21 & -   \\
Me 2-2   & 1.51 & 216  & -    & 4550 & 67.6 & 318  & 4.30\\
Vy 1-2   & -    & 163  & -    & -    & -    & -    & -   \\
\hline
\end{tabular}
\end{small}
\end{table}

In calculating the values of Table \ref{t:ion}, de-reddened fluxes have been
used according to \cite{2011EAS....47..189A}.
In addition to this, excitation class of each PNe was also calculated according
to \cite{1991Ap&SS.181...73G} in which the nebular lines have been used as a
(loosely) measure of the temperature of the central white dwarf.
Note that, if the stellar nuclei of the PNe is found to be very hot then it can
be taken as a \textit{high excitation PNe}.
Therefore, Hu 1-1, BoBn 1 and Vy 1-2 fall into this category.
However, if some related lines are missing from the central region then a
definitive calculation could not be made.
Therefore the following PNe fall into this \textit{unidentified} category:
M 1-4, K3-70, M1-17 SaSt 2-3 DdDm 1 M3-27 IC 4997 and Me 2-2.
Finally, the rest of the PNe can safely be classified as
\textit{medium excitation PNe}.

These studies can significantly contribute to the study of the evolution of the PNe and the Galaxy.
\section{Conclusions}
\label{sec:conc}

\begin{itemize}
\item
A similar study of D97 (for Southern Hemisphere) and W05 (for Northern Hemisphere) have been carried out by extending the northern hemisphere coverage of PNe.

\item
We present new emission line fluxes to be used as a standard in both imaging in narrow band and in Fabry-Perot spectroscopy.

\item
Emission line flux measurements of 12 PNe were made for the first time.

\item
Physical and chemical properties of PNe, as well as their evolution, can be studied with continuous monitoring of these 17 PNe.

\item
For suitable PNe, Extinction constant, Electron Temperature, Electron Density, Chemical Abundance and Excitation Class have been calculated.
\end{itemize}
\begin{acknowledgements}

The authors thank to the scientific and technological research council of Turkey (TUBITAK) for a partial support in using RTT150 (Russian-Turkish 1.5-m telescope) in Antalya through with project numbers 10BRTT150-33-0 and 10ARTT150-489-0. NA also thank to TUBITAK National Observatory (TUG) staff. NA gratefully acknowledges support through a Post-Doc Fellowship from the TUBITAK-BIDEB at Physics Department of Middle East Technical University, Ankara-Turkey. The author also grateful to M. E. Ozel for reading and correcting the manuscript and for his valuable remarks.

\end{acknowledgements}

\begin{table*}
 \caption{BoBn 1 fluxes and equivalent widths.}
 \label{tab:3}
\begin{small}
\begin{center}
\begin{tabular}{@{}c@{~}c@{~}c@{~}c@{~}c@{~}c@{~}c@{}}
\hline \hline
 & &\multicolumn{2}{c}{This Work} & & \multicolumn{2}{c}{W05}\\
\cline{3-4}
\cline{6-7}
  \mcc{Wavelength}
 & \mcc{Ion}
 & \mcc{log [Flux]}
 & \mcc{log [EW]}
 & \mcc{}
 & \mcc{log [Flux]}
 & \mcc{log [EW]}
  \\
  \mcc{{\AA}}
 & \mcc{}
 & \mcc{(ergs cm$^{-2}$ s$^{-1}$)}
 & \mcc{({\AA})}
 & \mcc{}
 & \mcc{(ergs cm$^{-2}$ s$^{-1}$)}
 & \mcc{({\AA})}
  \\
\hline
3967.41&[NeIII]&-13.680$\pm$0.008&2.719$\pm$0.240&&-&-\\
4101.74&H${\delta}$&-13.947$\pm$0.020&2.294$\pm$0.177&&-&-\\
4340.47&H${\gamma}$&-13.562$\pm$0.012&2.656$\pm$0.273&&-&-\\
4363.21&[OIII]&-14.356$\pm$0.155&1.507$\pm$0.647&&-&-\\
4387.93&HeI&-14.368$\pm$0.014&1.830$\pm$0.216&&-&-\\
4471.68&HeI&-14.195$\pm$0.013&2.264$\pm$0.066&&-&-\\
4685.71&HeII&-13.891$\pm$0.019&2.124$\pm$0.222&&-13.103$\pm$0.024&2.097$\pm$0.107\\
4861.20&H${\beta}$&-13.021$\pm$0.007&2.917$\pm$0.124&&-12.425$\pm$0.011&2.759$\pm$0.128\\
4958.52&[OIII]&-12.924$\pm$0.006&3.154$\pm$0.477&&-12.358$\pm$0.005&2.797$\pm$0.065\\
5754.59&[NII]&-&-&&-14.447$\pm$0.005&1.000$\pm$0.105\\
5007.57&[OIII]&-12.432$\pm$0.001&4.020$\pm$0.480&&-11.877$\pm$0.006&3.188$\pm$0.140\\
5875.97&HeI&-13.706$\pm$0.023&2.316$\pm$0.320&&-13.278$\pm$0.019&2.009$\pm$0.255\\
6300.30&[OI]&-&-&&-14.503$\pm$0.069&0.986$\pm$0.050\\
6548.05&[NII]&-13.344$\pm$0.716&2.010$\pm$0.392&&-13.296$\pm$0.014&1.674$\pm$0.050\\
6562.85&H${\alpha}$&-12.765$\pm$0.443&2.226$\pm$0.330&&-11.934$\pm$0.002&2.959$\pm$0.045\\
6583.45&[NII]&-13.351$\pm$0.011&2.449$\pm$0.181&&-12.828$\pm$0.012&2.166$\pm$0.059\\
6678.15&HeI&-14.504$\pm$0.191&1.413$\pm$0.400&&-13.853$\pm$0.022&1.585$\pm$0.206\\
7065.71&HeI&-&-&&-13.723$\pm$0.038&1.767$\pm$0.109\\
7135.80&[ArIII]&-14.161$\pm$0.072&1.699$\pm$0.354&&-&-\\
 \hline
 \end{tabular}
\end{center}
\end{small}
\end{table*}

\begin{table*}
 \caption{M 1-5 fluxes and equivalent widths.}
 \label{tab:4}
\begin{small}
\begin{center}
\begin{tabular}{@{}c@{~}c@{~}c@{~}c@{~}c@{~}c@{~}c@{}}
\hline \hline
 & &\multicolumn{2}{c}{This Work} & & \multicolumn{2}{c}{W05}\\
\cline{3-4}
\cline{6-7}
  \mcc{Wavelength}
 & \mcc{Ion}
 & \mcc{log [Flux]}
 & \mcc{log [EW]}
 & \mcc{}
 & \mcc{log [Flux]}
 & \mcc{log [EW]}
  \\
  \mcc{{\AA}}
 & \mcc{}
 & \mcc{(ergs cm$^{-2}$ s$^{-1}$)}
 & \mcc{({\AA})}
 & \mcc{}
 & \mcc{(ergs cm$^{-2}$ s$^{-1}$)}
 & \mcc{({\AA})}
  \\
\hline
4101.74&H${\delta}$&-14.201$\pm$0.378&1.399$\pm$0.817&&-&-\\
4340.47&H${\gamma}$&-14.199$\pm$0.000&2.263$\pm$0.079&&-&-\\
4685.71&HeII&-13.681$\pm$0.004&2.332$\pm$0.207&&-&-\\
4861.20&H${\beta}$&-13.559$\pm$0.004&2.496$\pm$0.035&&-&-\\
4958.52&[OIII]&-13.413$\pm$0.002&2.499$\pm$0.001&&-&-\\
5007.57&[OIII]&-12.901$\pm$0.001&3.081$\pm$0.072&&-&-\\
5269.20&[KVI]&-14.328$\pm$0.009&1.639$\pm$0.057&&-&-\\
5754.59&[NII]&-14.555$\pm$0.035&1.385$\pm$0.122&&-13.355$\pm$0.098&1.228$\pm$0.102\\
5875.97&HeI&-13.960$\pm$0.008&1.937$\pm$0.064&&-12.551$\pm$0.007&2.029$\pm$0.062\\
6300.30&[OI]&-14.795$\pm$0.148&1.128$\pm$0.332&&-13.273$\pm$0.032&1.265$\pm$0.038\\
6312.10&[SIII]&-&-&&-13.685$\pm$0.034&0.819$\pm$0.048\\
6548.05&[NII]&-15.115$\pm$0.080&0.939$\pm$0.118&&-12.086$\pm$0.002&1.793$\pm$0.076\\
6562.85&H${\alpha}$&-12.703$\pm$0.013&3.018$\pm$0.063&&-11.082$\pm$0.002&2.817$\pm$0.029\\
6583.45&[NII]&-13.216$\pm$0.018&2.510$\pm$0.026&&-11.595$\pm$0.001&2.498$\pm$0.063\\
6678.15&HeI&-14.519$\pm$0.080&1.273$\pm$0.054&&-12.940$\pm$0.001&1.572$\pm$0.032\\
6716.47&[SII]&-&-&&-13.497$\pm$0.005&1.008$\pm$0.026\\
6730.85&[SII]&-14.730$\pm$0.028&1.22$\pm$0.058&&-13.157$\pm$0.001&1.332$\pm$0.016\\
7065.71&HeI&-14.147$\pm$0.004&1.710$\pm$0.031&&-12.513$\pm$0.002&1.942$\pm$0.031\\
7135.80&[ArIII]&-13.785$\pm$0.607&2.115$\pm$0.584&&-12.530$\pm$0.010&1.919$\pm$0.045\\
 \hline
 \end{tabular}
\end{center}
\end{small}
\end{table*}

\begin{table*}
 \caption{NGC6833 fluxes and equivalent widths.}
 \label{tab:5}
\begin{small}
\begin{center}
\begin{tabular}{@{}c@{~}c@{~}c@{~}c@{~}c@{~}c@{~}c@{}}
\hline \hline
 & &\multicolumn{2}{c}{This Work} & & \multicolumn{2}{c}{W05}\\
\cline{3-4}
\cline{6-7}
  \mcc{Wavelength}
 & \mcc{Ion}
 & \mcc{log [Flux]}
 & \mcc{log [EW]}
 & \mcc{}
 & \mcc{log [Flux]}
 & \mcc{log [EW]}
  \\
  \mcc{{\AA}}
 & \mcc{}
 & \mcc{(ergs cm$^{-2}$ s$^{-1}$)}
 & \mcc{({\AA})}
 & \mcc{}
 & \mcc{(ergs cm$^{-2}$ s$^{-1}$)}
 & \mcc{({\AA})}
  \\
\hline
3697.15&H17&-&-&&-13.829$\pm$0.241&0.000$\pm$0.109\\
3705.00&HeI&-&-&&-13.288$\pm$0.081&0.423$\pm$0.037\\
3712.75&HeII&-&-&&-13.346$\pm$0.149&0.401$\pm$0.043\\
3726.19&[OII]&-&-&&-12.259$\pm$0.066&1.358$\pm$0.012\\
3734.37&H13&-&-&&-12.977$\pm$0.116&0.562$\pm$0.024\\
3750.15&H12&-&-&&-12.920$\pm$0.023&0.959$\pm$0.004\\
3770.63&H11&-&-&&-12.812$\pm$0.024&1.066$\pm$0.005\\
3797.90&H10&-&-&&-12.689$\pm$0.055&1.184$\pm$0.003\\
3819.70&HeI*&-&-&&-13.305$\pm$0.146&0.573$\pm$0.076\\
3835.38&H9&-&-&&-12.540$\pm$0.030&1.334$\pm$0.006\\
3868.71&[NeIII]&-&-&&-11.439$\pm$0.020&2.394$\pm$0.033\\
3889.05&H8&-12.771$\pm$0.004&2.219$\pm$0.043&&-12.142$\pm$0.019&1.692$\pm$0.024\\
3967.41&[NeIII]&-13.889$\pm$0.096&1.049$\pm$0.159&&-11.761$\pm$0.001&2.136$\pm$0.005\\
4026.10&HeI*&-12.793$\pm$0.011&2.123$\pm$0.104&&-13.025$\pm$0.013&0.905$\pm$0.014\\
4068.91&CIII&-&-&&-13.442$\pm$0.128&0.481$\pm$0.117\\
4101.74&H${\beta}$&-&-&&-11.954$\pm$0.006&1.988$\pm$0.001\\
4340.47&H${\gamma}$&-12.947$\pm$0.004&1.788$\pm$0.020&&-11.673$\pm$0.037&2.269$\pm$0.026\\
4363.21&[OIII]&-14.164$\pm$0.110&0.567$\pm$0.127&&-12.247$\pm$0.040&1.715$\pm$0.056\\
4387.93&HeI&-&-&&-13.601$\pm$0.042&0.378$\pm$0.006\\
4471.68&HeI&-&-&&-12.613$\pm$0.039&1.369$\pm$0.030\\
4685.71&HeII&-13.699$\pm$0.070&1.018$\pm$0.107&&-&-\\
4713.38&HeI*&-13.727$\pm$0.070&0.975$\pm$0.107&&-13.244$\pm$0.009&0.821$\pm$0.012\\
4724.30&[NeIV]&-13.884$\pm$0.144&0.828$\pm$0.188&&-&-\\
4740.18&[ArIV]&-&-&&-13.447$\pm$0.034&0.590$\pm$0.039\\
4861.20&H${\beta}$&-11.838$\pm$0.002&2.827$\pm$0.079&&-11.315$\pm$0.012&2.733$\pm$0.033\\
4921.93&HeI&-13.587$\pm$0.095&1.057$\pm$0.155&&-13.246$\pm$0.034&0.785$\pm$0.050\\
4958.52&[OIII]&-11.410$\pm$0.001&3.171$\pm$0.077&&-10.886$\pm$0.007&3.003$\pm$0.023\\
5007.57&[OIII]&-11.075$\pm$0.000&3.397$\pm$0.074&&-10.401$\pm$0.001&3.439$\pm$0.074\\
5191.80&[ArIII]&-&-&&-14.275$\pm$0.177&-0.150$\pm$0.186\\
5269.20&[KVI]&-13.314$\pm$0.024&1.219$\pm$0.036&&-&-\\
5537.89&[CIIII]&-&-&&-13.935$\pm$0.082&0.243$\pm$0.098\\
5754.59&[NII]&-13.314$\pm$0.024&1.219$\pm$0.036&&-13.049$\pm$0.047&1.121$\pm$0.017\\
5875.97&HeI&-12.423$\pm$0.001&2.129$\pm$0.011&&-12.038$\pm$0.002&2.112$\pm$0.011\\
6300.30&[OI]&-13.187$\pm$0.023&1.524$\pm$0.079&&-12.841$\pm$0.268&1.342$\pm$0.272\\
6312.10&[SIII]&-13.671$\pm$0.037&1.113$\pm$0.068&&-&-\\
6363.77&[OI]&-13.622$\pm$0.044&1.122$\pm$0.051&&-13.245$\pm$0.006&0.966$\pm$0.080\\
6548.05&[NII]&-12.723$\pm$0.030&1.870$\pm$0.010&&-12.262$\pm$0.007&1.247$\pm$0.038\\
6562.85&H${\alpha}$&-11.371$\pm$0.001&3.320$\pm$0.144&&-10.752$\pm$0.001&2.838$\pm$0.051\\
6583.45&[NII]&-12.239$\pm$0.001&2.367$\pm$0.013&&-11.865$\pm$0.003&1.867$\pm$0.021\\
6678.15&HeI&-13.048$\pm$0.002&1.661$\pm$0.010&&-12.620$\pm$0.007&1.598$\pm$0.028\\
6716.47&[SII]&-14.203$\pm$0.075&0.533$\pm$0.189&&-13.888$\pm$0.033&0.329$\pm$0.035\\
6730.85&[SII]&-13.897$\pm$0.111&0.864$\pm$0.074&&-13.531$\pm$0.029&0.700$\pm$0.001\\
7065.71&HeI&-12.572$\pm$0.001&2.170$\pm$0.011&&-12.130$\pm$0.003&2.123$\pm$0.009\\
7135.80&[ArIII]&-12.547$\pm$0.005&2.151$\pm$0.067&&-&-\\
 \hline
 \end{tabular}
\end{center}
\end{small}
\end{table*}

\begin{table*}
 \caption{IC5117 fluxes and equivalent widths.}
 \label{tab:6}
\begin{small}
\begin{center}
\begin{tabular}{@{}c@{~}c@{~}c@{~}c@{~}c@{~}c@{~}c@{}}
\hline \hline
 & &\multicolumn{2}{c}{This Work} & & \multicolumn{2}{c}{W05}\\
\cline{3-4}
\cline{6-7}
  \mcc{Wavelength}
 & \mcc{Ion}
 & \mcc{log [Flux]}
 & \mcc{log [EW]}
 & \mcc{}
 & \mcc{log [Flux]}
 & \mcc{log [EW]}
  \\
  \mcc{{\AA}}
 & \mcc{}
 & \mcc{(ergs cm$^{-2}$ s$^{-1}$)}
 & \mcc{({\AA})}
 & \mcc{}
 & \mcc{(ergs cm$^{-2}$ s$^{-1}$)}
 & \mcc{({\AA})}
  \\
\hline
3705.00&HeI&-&-&&-13.661$\pm$0.155&0.810$\pm$0.211\\
3712.75&HeII&-&-&&-13.656$\pm$0.032&0.806$\pm$0.122\\
3726.19&[OII]&-&-&&-12.337$\pm$0.004&1.768$\pm$0.023\\
3734.37&H13&-&-&&-13.340$\pm$0.088&0.493$\pm$0.261\\
3750.15&H12&-&-&&-13.254$\pm$0.021&1.217$\pm$0.179\\
3770.63&H11&-&-&&-13.156$\pm$0.053&1.502$\pm$0.051\\
3797.90&H10&-&-&&-13.014$\pm$0.005&1.651$\pm$0.175\\
3819.70&HeI*&-&-&&-13.700$\pm$0.001&0.846$\pm$0.019\\
3835.38&H9&-&-&&-12.836$\pm$0.015&1.767$\pm$0.016\\
3868.71&[NeIII]&-&-&&-11.503$\pm$0.002&2.871$\pm$0.071\\
3889.05&H8&-&-&&-12.482$\pm$0.013&1.955$\pm$0.098\\
3967.41&[NeIII]&-&-&&-11.844$\pm$0.003&2.781$\pm$0.100\\
4068.91&CIII&-&-&&-12.991$\pm$0.002&1.382$\pm$0.063\\
4101.74&H${\delta}$&-12.943$\pm$0.214&2.696$\pm$0.214&&-12.179$\pm$0.012&2.387$\pm$0.134\\
4340.47&H${\gamma}$&-12.537$\pm$0.388&2.845$\pm$0.388&&-11.876$\pm$0.003&2.581$\pm$0.112\\
4363.21&[OIII]&-12.832$\pm$0.328&2.752$\pm$0.328&&-12.191$\pm$0.010&2.262$\pm$0.077\\
4471.68&HeI&-13.365$\pm$0.265&2.288$\pm$0.265&&-12.798$\pm$0.001&1.783$\pm$0.024\\
4634.14&NIII&-13.542$\pm$0.624&1.209$\pm$0.624&&-13.576$\pm$0.036&0.695$\pm$0.333\\
4640.64&NIII&-&-&&-13.215$\pm$0.013&0.947$\pm$0.291\\
4685.71&HeII&-12.888$\pm$0.464&2.473$\pm$0.464&&-12.359$\pm$0.001&2.106$\pm$0.075\\
4713.38&HeI*&-13.555$\pm$0.549&1.765$\pm$0.549&&-13.036$\pm$0.003&1.419$\pm$0.082\\
4724.30&[NeIV]&-&-&&-14.154$\pm$0.057&0.270$\pm$0.163\\
4740.18&[ArIV]&-13.175$\pm$0.206&2.047$\pm$0.206&&-12.661$\pm$0.001&1.813$\pm$0.029\\
4861.20&H${\beta}$&-11.873$\pm$0.212&3.331$\pm$0.212&&-11.380$\pm$0.020&3.032$\pm$0.006\\
4921.93&HeI&-13.644$\pm$0.448&1.442$\pm$0.448&&-13.180$\pm$0.050&1.019$\pm$0.063\\
4958.52&[OIII]&-11.093$\pm$0.187&3.687$\pm$0.187&&-10.629$\pm$0.023&3.270$\pm$0.124\\
5007.57&[OIII]&-10.646$\pm$0.294&3.560$\pm$0.294&&-10.131$\pm$0.017&3.743$\pm$0.026\\
5191.80&[ArIII]&-&-&&-14.152$\pm$0.159&0.151$\pm$0.004\\
5197.90&[NI]&-&-&&-13.852$\pm$0.019&0.488$\pm$0.116\\
5411.52&HeII&-13.519$\pm$0.094&1.361$\pm$0.094&&-13.276$\pm$0.042&1.107$\pm$0.019\\
5517.72&[CIIII]&-&-&&-14.084$\pm$0.025&0.209$\pm$0.072\\
5537.89&[CIIII]&-13.635$\pm$0.005&1.138$\pm$0.005&&-13.544$\pm$0.005&0.801$\pm$0.007\\
5754.59&[NII]&-12.931$\pm$0.048&1.798$\pm$0.048&&-12.653$\pm$0.006&1.650$\pm$0.005\\
5875.97&HeI&-&-&&-11.909$\pm$0.024&2.325$\pm$0.109\\
6300.30&[OI]&-12.638$\pm$0.056&2.155$\pm$0.056&&-12.177$\pm$0.026&1.881$\pm$0.050\\
6312.10&[SIII]&-13.010$\pm$0.057&1.868$\pm$0.057&&-12.606$\pm$0.039&1.437$\pm$0.011\\
6363.77&[OI]&-13.073$\pm$0.153&1.756$\pm$0.153&&-12.656$\pm$0.024&1.453$\pm$0.038\\
6548.05&[NII]&-12.245$\pm$0.119&2.398$\pm$0.119&&-11.733$\pm$0.032&1.612$\pm$0.116\\
6562.85&H${\alpha}$&-11.024$\pm$0.162&3.588$\pm$0.162&&-10.524$\pm$0.002&2.887$\pm$0.223\\
6583.45&[NII]&-11.781$\pm$0.117&2.797$\pm$0.117&&-11.272$\pm$0.008&2.312$\pm$0.184\\
6678.15&HeI&-12.850$\pm$0.008&1.832$\pm$0.008&&-12.366$\pm$0.003&1.743$\pm$0.007\\
6716.47&[SII]&-13.448$\pm$0.042&1.214$\pm$0.042&&-12.958$\pm$0.006&1.151$\pm$0.015\\
6730.85&[SII]&-13.156$\pm$0.155&1.693$\pm$0.155&&-12.612$\pm$0.008&1.495$\pm$0.037\\
7065.71&HeI&-12.325$\pm$0.399&2.461$\pm$0.399&&-11.860$\pm$0.009&2.200$\pm$0.044\\
7135.80&[ArIII]&-12.097$\pm$0.114&2.465$\pm$0.114&&-&-\\
 \hline
 \end{tabular}
\end{center}
\end{small}
\end{table*}

\begin{table*}
 \caption{Hu 1-1 fluxes and equivalent widths.}
 \label{tab:7}
\begin{small}
\begin{center}
\begin{tabular}{@{}c@{~}c@{~}c@{~}c@{}}
\hline \hline
   \mcc{Wavelength}
 & \mcc{Ion}
 & \mcc{log [Flux]}
 & \mcc{log [EW]}
  \\
  \mcc{{\AA}}
 & \mcc{}
 & \mcc{(ergs cm$^{-2}$ s$^{-1}$)}
 & \mcc{({\AA})}
  \\
\hline
3726.19&[OII]&-11.424$\pm$0.002&2.967$\pm$0.018\\
3797.90&H10&-13.353$\pm$0.011&1.217$\pm$0.006\\
3835.38&H9&-13.065$\pm$0.006&1.519$\pm$0.005\\
3868.71&[NeIII]&-11.768$\pm$0.005&2.776$\pm$0.094\\
3889.05&H8&-12.524$\pm$0.023&2.160$\pm$0.142\\
3967.41&[NeIII]&-12.108$\pm$0.011&2.727$\pm$0.239\\
4026.10&HeI*&-13.547$\pm$0.057&1.258$\pm$0.007\\
4068.91&CIII&-12.929$\pm$0.020&1.848$\pm$0.064\\
4101.74&H${\delta}$&-12.416$\pm$0.008&2.352$\pm$0.065\\
4340.47&H${\gamma}$&-12.163$\pm$0.012&2.607$\pm$0.168\\
4363.21&[OIII]&-12.743$\pm$0.044&2.027$\pm$0.200\\
4471.68&HeI&-13.010$\pm$0.084&1.676$\pm$0.213\\
4640.64&NIII&-13.518$\pm$0.080&1.187$\pm$0.058\\
4685.71&HeII&-12.498$\pm$0.035&2.677$\pm$0.695\\
4713.38&HeI*&-13.574$\pm$0.064&2.469$\pm$0.048\\
4861.20&H${\beta}$&-11.793$\pm$0.001&2.953$\pm$0.035\\
4921.93&HeI&-13.408$\pm$0.040&1.334$\pm$0.068\\
4959.52&[OIII]&-11.179$\pm$0.002&3.500$\pm$0.186\\
5007.57&[OIII]&-10.734$\pm$0.001&3.811$\pm$0.158\\
5197.90&[NI]&-13.375$\pm$0.159&1.479$\pm$0.181\\
5411.52&HeII&-13.449$\pm$0.074&1.337$\pm$0.099\\
5537.89&[CIIII]&-13.109$\pm$0.063&1.910$\pm$0.286\\
5875.97&HeI&-12.539$\pm$0.005&2.346$\pm$0.046\\
6300.30&[OI]&-12.378$\pm$0.008&2.540$\pm$0.014\\
6363.77&[OI]&-12.856$\pm$0.008&2.101$\pm$0.039\\
6562.85&H${\alpha}$&-10.912$\pm$0.001&1.583$\pm$0.002\\
6678.15&HeI&-13.126$\pm$0.019&1.763$\pm$0.061\\
6730.85&[SII]&-11.911$\pm$0.002&2.972$\pm$0.050\\
7065.71&HeI&-13.104$\pm$0.011&1.826$\pm$0.038\\
7135.80&[ArIII]&-12.384$\pm$0.003&2.550$\pm$0.037\\
 \hline
 \end{tabular}
\end{center}
\end{small}
\end{table*}

\begin{table*}
 \caption{M 1-4 fluxes and equivalent widths.}
 \label{tab:8}
\begin{small}
\begin{center}
\begin{tabular}{@{}c@{~}c@{~}c@{~}c@{}}
\hline \hline
   \mcc{Wavelength}
 & \mcc{Ion}
 & \mcc{log [Flux]}
 & \mcc{log [EW]}
  \\
  \mcc{{\AA}}
 & \mcc{}
 & \mcc{(ergs cm$^{-2}$ s$^{-1}$)}
 & \mcc{({\AA})}
  \\
\hline
4861.20&H${\beta}$&-12.610$\pm$0.087&1.350$\pm$0.183\\
4959.52&[OIII]&-11.942$\pm$0.021&1.930$\pm$0.104\\
5007.57&[OIII]&-11.462$\pm$0.007&2.456$\pm$0.108\\
5875.97&HeI&-12.845$\pm$0.003&1.883$\pm$0.013\\
6300.30&[OI]&-14.304$\pm$0.056&0.469$\pm$0.085\\
6312.10&[SIII]&-13.670$\pm$0.011&0.469$\pm$0.085\\
6562.85&H${\alpha}$&-11.443$\pm$0.001&3.130$\pm$0.056\\
6583.45&[NII]&-12.854$\pm$0.022&1.729$\pm$0.071\\
6678.15&HeI&-13.292$\pm$0.048&1.233$\pm$0.086\\
6716.47&[SII]&-13.878$\pm$0.015&0.665$\pm$0.003\\
6730.85&[SII]&-13.810$\pm$0.304&0.714$\pm$0.378\\
7065.71&HeI&-12.973$\pm$0.054&1.519$\pm$0.134\\
7135.80&[ArIII]&-12.820$\pm$0.020&1.670$\pm$0.063\\
 \hline
 \end{tabular}
\end{center}
\end{small}
\end{table*}

\begin{table*}
 \caption{K 3-70 fluxes and equivalent widths.}
 \label{tab:9}
\begin{small}
\begin{center}
\begin{tabular}{@{}c@{~}c@{~}c@{~}c@{}}
\hline \hline
   \mcc{Wavelength}
 & \mcc{Ion}
 & \mcc{log [Flux]}
 & \mcc{log [EW]}
  \\
  \mcc{{\AA}}
 & \mcc{}
 & \mcc{(ergs cm$^{-2}$ s$^{-1}$)}
 & \mcc{({\AA})}
  \\
\hline
4959.52&[OIII]&-13.546$\pm$0.005&2.871$\pm$0.239\\
5007.57&[OIII]&-13.045$\pm$0.001&3.276$\pm$0.164\\
5754.59&[NII]&-14.256$\pm$0.042&2.030$\pm$0.309\\
5875.97&HeI&-14.360$\pm$0.009&2.125$\pm$0.147\\
6300.30&[OI]&-14.815$\pm$0.087&1.648$\pm$0.135\\
6312.10&[SIII]&-14.994$\pm$0.094&1.958$\pm$0.653\\
6548.05&[NII]&-13.204$\pm$0.001&3.465$\pm$0.080\\
6562.85&H${\alpha}$&-13.130$\pm$0.004&3.046$\pm$0.289\\
6583.45&[NII]&-12.720$\pm$0.002&3.306$\pm$0.245\\
6678.15&HeI&-14.903$\pm$0.053&1.932$\pm$0.486\\
6716.47&[SII]&-14.440$\pm$0.051&1.858$\pm$0.258\\
6730.85&[SII]&-14.293$\pm$0.005&2.160$\pm$0.077\\
7065.71&HeI&-14.555$\pm$0.045&2.090$\pm$0.341\\
7135.80&[ArIII]&-14.139$\pm$0.006&2.515$\pm$0.204\\
 \hline
 \end{tabular}
\end{center}
\end{small}
\end{table*}

\begin{table*}
 \caption{M 1-17 fluxes and equivalent widths.}
 \label{tab:10}
\begin{small}
\begin{center}
\begin{tabular}{@{}c@{~}c@{~}c@{~}c@{}}
\hline \hline
   \mcc{Wavelength}
 & \mcc{Ion}
 & \mcc{log [Flux]}
 & \mcc{log [EW]}
  \\
  \mcc{{\AA}}
 & \mcc{}
 & \mcc{(ergs cm$^{-2}$ s$^{-1}$)}
 & \mcc{({\AA})}
  \\
\hline
4861.20&H${\beta}$&-12.734$\pm$0.004&3.201$\pm$0.391\\
4959.52&[OIII]&-12.014$\pm$0.001&3.751$\pm$0.415\\
5007.57&[OIII]&-11.511$\pm$0.000&4.007$\pm$0.211\\
5197.90&[NI]&-14.116$\pm$0.098&1.748$\pm$0.371\\
5754.59&[NII]&-13.645$\pm$0.116&2.026$\pm$0.662\\
5875.97&HeI&-13.218$\pm$0.013&2.294$\pm$0.174\\
6300.30&[OI]&-13.177$\pm$0.005&2.361$\pm$0.027\\
6312.10&[SIII]&-13.948$\pm$0.057&1.609$\pm$0.244\\
6363.77&[OI]&-13.601$\pm$0.001&1.940$\pm$0.001\\
6548.05&[NII]&-12.529$\pm$0.000&2.919$\pm$0.002\\
6562.85&H${\alpha}$&-11.948$\pm$0.000&3.462$\pm$0.001\\
6583.45&[NII]&-12.101$\pm$0.004&3.361$\pm$0.236\\
6678.15&HeI&-13.813$\pm$0.009&1.835$\pm$0.460\\
6716.47&[SII]&-13.363$\pm$0.004&2.248$\pm$0.076\\
6730.85&[SII]&-13.144$\pm$0.001&2.486$\pm$0.044\\
7065.71&HeI&-13.558$\pm$0.011&1.909$\pm$0.092\\
7135.80&[ArIII]&-12.968$\pm$0.000&2.516$\pm$0.002\\
 \hline
 \end{tabular}
\end{center}
\end{small}
\end{table*}

\begin{table*}
 \caption{SaSt 2-3 fluxes and equivalent widths.}
 \label{tab:11}
\begin{small}
\begin{center}
\begin{tabular}{@{}c@{~}c@{~}c@{~}c@{}}
\hline \hline
   \mcc{Wavelength}
 & \mcc{Ion}
 & \mcc{log [Flux]}
 & \mcc{log [EW]}
  \\
  \mcc{{\AA}}
 & \mcc{}
 & \mcc{(ergs cm$^{-2}$ s$^{-1}$)}
 & \mcc{({\AA})}
  \\
\hline
4068.91&CIII&-13.812$\pm$0.027&0.701$\pm$0.035\\
4713.38&HeI*&-14.145$\pm$0.018&0.191$\pm$0.022\\
4861.20&H${\beta}$&-12.694$\pm$0.001&1.599$\pm$0.002\\
5007.57&[OIII]&-14.115$\pm$0.002&0.166$\pm$0.002\\
5754.59&[NII]&-14.241$\pm$0.047&0.007$\pm$0.001\\
6300.30&[OI]&-13.715$\pm$0.007&0.748$\pm$0.015\\
6548.05&[NII]&-12.942$\pm$0.002&1.615$\pm$0.009\\
6562.85&H${\alpha}$&-12.114$\pm$0.000&2.534$\pm$0.003\\
6583.45&[NII]&-12.553$\pm$0.001&2.093$\pm$0.007\\
6716.47&[SII]&-13.900$\pm$0.008&0.684$\pm$0.012\\
6730.85&[SII]&-13.723$\pm$0.007&0.863$\pm$0.008\\
7135.80&[ArIII]&-14.719$\pm$0.007&0.783$\pm$0.016\\
 \hline
 \end{tabular}
\end{center}
\end{small}
\end{table*}

\begin{table*}
 \caption{H 4-1 fluxes and equivalent widths.}
 \label{tab:12}
\begin{small}
\begin{center}
\begin{tabular}{@{}c@{~}c@{~}c@{~}c@{}}
\hline \hline
   \mcc{Wavelength}
 & \mcc{Ion}
 & \mcc{log [Flux]}
 & \mcc{log [EW]}
  \\
  \mcc{{\AA}}
 & \mcc{}
 & \mcc{(ergs cm$^{-2}$ s$^{-1}$)}
 & \mcc{({\AA})}
  \\
\hline
3712.75&HeII&-12.403$\pm$0.004&2.687$\pm$0.093\\
3734.37&H13&-14.359$\pm$0.052&0.837$\pm$0.053\\
3750.15&H12&-14.125$\pm$0.067&1.170$\pm$0.158\\
3770.63&H11&-13.918$\pm$0.004&1.439$\pm$0.007\\
3819.70&HeI*&-13.731$\pm$0.016&1.652$\pm$0.049\\
3835.38&H9&-13.823$\pm$0.008&1.545$\pm$0.020\\
3868.71&[NeIII]&-13.226$\pm$0.000&2.132$\pm$0.003\\
4068.91&CIII&-13.181$\pm$0.010&2.329$\pm$0.122\\
4340.47&H${\gamma}$&-12.954$\pm$0.003&2.537$\pm$0.060\\
4363.21&[OIII]&-13.698$\pm$0.025&1.774$\pm$0.109\\
4387.93&HeI&-14.709$\pm$0.074&0.809$\pm$0.093\\
4471.68&HeI&-13.887$\pm$0.042&1.583$\pm$0.127\\
4685.71&HeII&-13.567$\pm$0.011&2.079$\pm$0.091\\
4861.20&H${\beta}$&-12.570$\pm$0.000&3.252$\pm$0.014\\
4921.93&HeI&-14.177$\pm$0.006&1.549$\pm$0.014\\
4959.52&[OIII]&-12.247$\pm$0.000&3.332$\pm$0.014\\
5007.57&[OIII]&-11.773$\pm$0.000&3.746$\pm$0.099\\
5197.90&[NI]&-14.359$\pm$0.052&1.272$\pm$0.022\\
5411.52&HeII&-14.525$\pm$0.067&1.049$\pm$0.017\\
5754.59&[NII]&-14.249$\pm$0.005&1.532$\pm$0.004\\
6300.30&[OI]&-14.290$\pm$0.065&1.308$\pm$0.041\\
6363.77&[OI]&-13.950$\pm$0.044&1.753$\pm$0.169\\
6548.05&[NII]&-13.018$\pm$0.153&2.026$\pm$0.937\\
6562.85&H${\alpha}$&-12.060$\pm$0.017&3.013$\pm$0.858\\
6583.45&[NII]&-12.452$\pm$0.010&2.921$\pm$0.439\\
6678.15&HeI&-13.875$\pm$0.038&1.946$\pm$0.226\\
6716.47&[SII]&-14.518$\pm$0.029&1.136$\pm$0.049\\
6730.85&[SII]&-14.534$\pm$0.049&1.119$\pm$0.099\\
7065.71&HeI&-13.820$\pm$0.011&1.869$\pm$0.064\\
7135.80&[ArIII]&-14.634$\pm$0.003&1.115$\pm$0.007\\
 \hline
 \end{tabular}
\end{center}
\end{small}
\end{table*}

\begin{table*}
 \caption{DdDm 1 fluxes and equivalent widths.}
 \label{tab:13}
\begin{small}
\begin{center}
\begin{tabular}{@{}c@{~}c@{~}c@{~}c@{}}
\hline \hline
   \mcc{Wavelength}
 & \mcc{Ion}
 & \mcc{log [Flux]}
 & \mcc{log [EW]}
  \\
  \mcc{{\AA}}
 & \mcc{}
 & \mcc{(ergs cm$^{-2}$ s$^{-1}$)}
 & \mcc{({\AA})}
  \\
\hline
5875.97&HeI&-14.549$\pm$0.151&2.163$\pm$0.247\\
6300.30&[OI]&-14.871$\pm$0.069&1.998$\pm$0.623\\
6312.10&[SIII]&-15.104$\pm$0.109&1.508$\pm$0.414\\
6548.05&[NII]&-13.346$\pm$0.002&3.233$\pm$0.303\\
6562.85&H${\alpha}$&-13.247$\pm$0.001&3.258$\pm$0.278\\
6583.45&[NII]&-12.847$\pm$0.000&3.668$\pm$0.229\\
6678.15&HeI&-15.090$\pm$0.027&1.697$\pm$0.192\\
6716.47&[SII]&-14.480$\pm$0.001&2.499$\pm$0.046\\
6730.85&[SII]&-14.331$\pm$0.006&2.458$\pm$0.265\\
7065.71&HeI&-14.736$\pm$0.017&2.693$\pm$0.841\\
7135.80&[ArIII]&-14.245$\pm$0.002&3.372$\pm$0.627\\
 \hline
 \end{tabular}
\end{center}
\end{small}
\end{table*}

\begin{table*}
 \caption{Na 1 fluxes and equivalent widths.}
 \label{tab:14}
\begin{small}
\begin{center}
\begin{tabular}{@{}c@{~}c@{~}c@{~}c@{}}
\hline \hline
   \mcc{Wavelength}
 & \mcc{Ion}
 & \mcc{log [Flux]}
 & \mcc{log [EW]}
  \\
  \mcc{{\AA}}
 & \mcc{}
 & \mcc{(ergs cm$^{-2}$ s$^{-1}$)}
 & \mcc{({\AA})}
  \\
\hline
3726.19&[OII]&-13.408$\pm$0.053&1.270$\pm$0.078\\
3868.71&[NeIII]&-12.329$\pm$0.004&2.348$\pm$0.038\\
3889.05&H8&-13.090$\pm$0.033&1.551$\pm$0.074\\
3967.41&[NeIII]&-12.681$\pm$0.011&2.081$\pm$0.066\\
4101.74&H${\delta}$&-12.908$\pm$0.017&1.895$\pm$0.060\\
4340.47&H${\gamma}$&-12.660$\pm$0.001&2.135$\pm$0.002\\
4363.21&[OIII]&-13.133$\pm$0.019&1.667$\pm$0.056\\
4471.68&HeI&-13.481$\pm$0.168&1.396$\pm$0.286\\
4640.64&NIII&-13.409$\pm$0.004&1.526$\pm$0.000\\
4685.71&HeII&-12.997$\pm$0.019&1.849$\pm$0.072\\
4713.38&HeI*&-13.479$\pm$0.076&1.353$\pm$0.141\\
4740.18&[ArIV]&-13.602$\pm$0.112&1.242$\pm$0.180\\
4861.20&H${\beta}$&-12.175$\pm$0.003&2.759$\pm$0.062\\
4959.52&[OIII]&-11.497$\pm$0.001&3.216$\pm$0.071\\
5007.57&[OIII]&-11.014$\pm$0.001&3.659$\pm$0.178\\
5197.90&[NI]&-14.205$\pm$0.520&0.708$\pm$0.585\\
5411.52&HeII&-13.913$\pm$0.005&0.920$\pm$0.006\\
5875.97&HeI&-12.692$\pm$0.001&2.113$\pm$0.006\\
6300.30&[OI]&-14.600$\pm$0.118&0.332$\pm$0.146\\
6312.10&[SIII]&-14.048$\pm$0.014&0.832$\pm$0.012\\
6562.85&H${\alpha}$&-11.419$\pm$0.000&3.417$\pm$0.024\\
6583.45&[NII]&-13.347$\pm$0.045&1.477$\pm$0.079\\
6678.15&HeI&-13.235$\pm$0.020&1.626$\pm$0.056\\
6730.85&[SII]&-13.748$\pm$0.061&1.136$\pm$0.085\\
7065.71&HeI&-13.227$\pm$0.004&1.639$\pm$0.013\\
7135.80&[ArIII]&-13.013$\pm$0.003&1.858$\pm$0.011\\
 \hline
 \end{tabular}
\end{center}
\end{small}
\end{table*}

\begin{table*}
 \caption{Vy 1-2 fluxes and equivalent widths.}
 \label{tab:15}
\begin{small}
\begin{center}
\begin{tabular}{@{}c@{~}c@{~}c@{~}c@{}}
\hline \hline
   \mcc{Wavelength}
 & \mcc{Ion}
 & \mcc{log [Flux]}
 & \mcc{log [EW]}
  \\
  \mcc{{\AA}}
 & \mcc{}
 & \mcc{(ergs cm$^{-2}$ s$^{-1}$)}
 & \mcc{({\AA})}
  \\
\hline
3750.15&H12&-12.118$\pm$0.011&1.832$\pm$0.043\\
3797.90&H10&-12.631$\pm$0.001&1.567$\pm$0.048\\
3819.70&HeI*&-12.895$\pm$0.036&1.420$\pm$0.095\\
3868.71&[NeIII]&-12.795$\pm$0.010&1.417$\pm$0.029\\
3889.05&H8&-11.578$\pm$0.003&2.655$\pm$0.014\\
4026.10&HeI*&-13.224$\pm$0.010&1.198$\pm$0.018\\
4068.91&CIII&-13.000$\pm$0.020&1.396$\pm$0.034\\
4101.74&H${\delta}$&-12.120$\pm$0.007&2.313$\pm$0.069\\
4340.47&H${\gamma}$&-11.928$\pm$0.002&2.496$\pm$0.025\\
4363.21&[OIII]&-12.638$\pm$0.007&1.766$\pm$0.013\\
4471.68&HeI&-12.938$\pm$0.060&1.551$\pm$0.162\\
4634.14&NIII&-12.767$\pm$0.069&1.721$\pm$0.171\\
4685.71&HeII&-12.101$\pm$0.008&2.341$\pm$0.107\\
4713.38&HeI*&-12.926$\pm$0.082&1.532$\pm$0.217\\
4740.18&[ArIV]&-12.992$\pm$0.061&1.497$\pm$0.163\\
4861.20&H${\beta}$&-11.552$\pm$0.005&2.981$\pm$0.215\\
4921.93&HeI&-12.860$\pm$0.007&1.754$\pm$0.003\\
4959.52&[OIII]&-11.052$\pm$0.003&3.369$\pm$0.241\\
5007.57&[OIII]&-10.700$\pm$0.002&3.711$\pm$0.253\\
5411.52&HeII&-13.088$\pm$0.019&1.679$\pm$0.055\\
5875.97&HeI&-12.326$\pm$0.014&2.466$\pm$0.160\\
6300.30&[OI]&-13.238$\pm$0.057&1.547$\pm$0.082\\
6312.10&[SIII]&-13.242$\pm$0.028&1.641$\pm$0.191\\
6363.77&[OI]&-13.685$\pm$0.016&1.080$\pm$0.027\\
6548.05&[NII]&-12.576$\pm$0.045&2.071$\pm$0.008\\
6562.85&H${\alpha}$&-11.201$\pm$0.000&3.496$\pm$0.121\\
6583.45&[NII]&-12.147$\pm$0.005&2.630$\pm$0.114\\
6678.15&HeI&-12.973$\pm$0.006&1.808$\pm$0.026\\
6716.47&[SII]&-13.163$\pm$0.023&1.634$\pm$0.066\\
6730.85&[SII]&-12.954$\pm$0.003&1.820$\pm$0.011\\
7065.71&HeI&-12.973$\pm$0.001&1.823$\pm$0.007\\
7135.80&[ArIII]&-12.394$\pm$0.002&2.384$\pm$0.021\\
 \hline
 \end{tabular}
\end{center}
\end{small}
\end{table*}

\begin{table*}
 \caption{M 3-27 fluxes and equivalent widths.}
 \label{tab:16}
\begin{small}
\begin{center}
\begin{tabular}{@{}c@{~}c@{~}c@{~}c@{}}
\hline \hline
   \mcc{Wavelength}
 & \mcc{Ion}
 & \mcc{log [Flux]}
 & \mcc{log [EW]}
  \\
  \mcc{{\AA}}
 & \mcc{}
 & \mcc{(ergs cm$^{-2}$ s$^{-1}$)}
 & \mcc{({\AA})}
  \\
\hline
3770.63&H11&-12.014$\pm$0.003&2.182$\pm$0.023\\
3797.90&H10&-12.964$\pm$0.038&1.222$\pm$0.066\\
3967.41&[NeIII]&-13.296$\pm$0.013&0.926$\pm$0.019\\
4026.10&HeI*&-13.161$\pm$0.055&1.090$\pm$0.076\\
4068.91&CIII&-12.760$\pm$0.014&1.479$\pm$0.033\\
4340.47&H${\gamma}$&-12.513$\pm$0.023&1.761$\pm$0.086\\
4363.21&[OIII]&-12.031$\pm$0.011&2.251$\pm$0.098\\
4471.68&HeI&-12.978$\pm$0.056&1.320$\pm$0.107\\
4713.38&HeI*&-13.521$\pm$0.138&0.753$\pm$0.168\\
4861.20&H${\beta}$&-11.930$\pm$0.004&2.353$\pm$0.041\\
4921.93&HeI&-13.377$\pm$0.040&0.946$\pm$0.054\\
5875.97&HeI&-12.186$\pm$0.014&2.164$\pm$0.146\\
6300.30&[OI]&-12.471$\pm$0.001&1.845$\pm$0.031\\
6312.10&[SIII]&-12.841$\pm$0.013&1.634$\pm$0.001\\
6363.77&[OI]&-12.710$\pm$0.035&1.685$\pm$0.117\\
6548.05&[NII]&-10.958$\pm$0.002&2.395$\pm$0.015\\
6562.85&H${\alpha}$&-10.607$\pm$0.001&3.417$\pm$0.160\\
6678.15&HeI&-12.887$\pm$0.049&1.328$\pm$0.121\\
6730.85&[SII]&-12.671$\pm$0.097&0.729$\pm$0.111\\
7065.71&HeI&-12.195$\pm$0.003&2.083$\pm$0.024\\
7135.80&[ArIII]&-12.997$\pm$0.047&1.339$\pm$0.117\\
 \hline
 \end{tabular}
\end{center}
\end{small}
\end{table*}

\begin{table*}
 \caption{IC 4997 fluxes and equivalent widths.}
 \label{tab:17}
\begin{small}
\begin{center}
\begin{tabular}{@{}c@{~}c@{~}c@{~}c@{}}
\hline \hline
   \mcc{Wavelength}
 & \mcc{Ion}
 & \mcc{log [Flux]}
 & \mcc{log [EW]}
  \\
  \mcc{{\AA}}
 & \mcc{}
 & \mcc{(ergs cm$^{-2}$ s$^{-1}$)}
 & \mcc{({\AA})}
  \\
\hline
3734.37&H13&-11.186$\pm$0.057&2.102$\pm$0.307\\
3770.63&H11&-12.230$\pm$0.080&1.051$\pm$0.161\\
3797.90&H10&-11.917$\pm$0.053&1.364$\pm$0.137\\
3819.70&HeI*&-11.788$\pm$0.026&1.508$\pm$0.085\\
3835.38&H9&-12.395$\pm$0.120&0.882$\pm$0.188\\
3868.71&[NeIII]&-10.466$\pm$0.002&2.813$\pm$0.079\\
3889.05&H8&-11.290$\pm$0.013&1.968$\pm$0.101\\
3967.41&[NeIII]&-10.790$\pm$0.002&2.450$\pm$0.039\\
4026.10&HeI*&-12.143$\pm$0.039&1.109$\pm$0.080\\
4068.91&CIII&-11.959$\pm$0.064&1.338$\pm$0.124\\
4101.74&H${\delta}$&-11.105$\pm$0.007&2.182$\pm$0.080\\
4340.47&H${\gamma}$&-10.878$\pm$0.001&2.419$\pm$0.028\\
4363.21&[OIII]&-10.822$\pm$0.004&2.468$\pm$0.095\\
4387.93&HeI&-12.580$\pm$0.010&0.872$\pm$0.001\\
4471.68&HeI&-11.770$\pm$0.015&1.603$\pm$0.060\\
4640.64&NIII&-12.296$\pm$0.071&1.120$\pm$0.003\\
4713.38&HeI*&-12.384$\pm$0.016&0.999$\pm$0.030\\
4740.18&[ArIV]&-12.571$\pm$0.274&0.883$\pm$0.366\\
4861.20&H${\beta}$&-10.587$\pm$0.002&2.742$\pm$0.066\\
4921.93&HeI&-12.309$\pm$0.057&1.049$\pm$0.100\\
4959.52&[OIII]&-10.342$\pm$0.002&2.959$\pm$0.090\\
5007.57&[OIII]&-9.947$\pm$0.002&3.306$\pm$0.190\\
5754.59&[NII]&-12.380$\pm$0.008&1.156$\pm$0.017\\
5875.97&HeI&-11.122$\pm$0.004&2.393$\pm$0.051\\
6300.30&[OI]&-11.638$\pm$0.011&1.874$\pm$0.067\\
6312.10&[SIII]&-11.937$\pm$0.010&1.561$\pm$0.063\\
6363.77&[OI]&-12.103$\pm$0.004&1.348$\pm$0.010\\
6548.05&[NII]&-11.451$\pm$0.010&1.912$\pm$0.069\\
6562.85&H${\alpha}$&-10.038$\pm$0.000&3.315$\pm$0.064\\
6583.45&[NII]&-11.050$\pm$0.006&2.294$\pm$0.087\\
6678.15&HeI&-11.714$\pm$0.008&1.749$\pm$0.049\\
6716.47&[SII]&-12.464$\pm$0.059&1.002$\pm$0.103\\
6730.85&[SII]&-12.176$\pm$0.007&1.273$\pm$0.017\\
7065.71&HeI&-11.229$\pm$0.002&2.239$\pm$0.037\\
7135.80&[ArIII]&-11.343$\pm$0.001&2.123$\pm$0.019\\
 \hline
 \end{tabular}
\end{center}
\end{small}
\end{table*}

\begin{table*}
 \caption{Me 2-2 fluxes and equivalent widths.}
 \label{tab:18}
\begin{small}
\begin{center}
\begin{tabular}{@{}c@{~}c@{~}c@{~}c@{}}
\hline \hline
   \mcc{Wavelength}
 & \mcc{Ion}
 & \mcc{log [Flux]}
 & \mcc{log [EW]}
  \\
  \mcc{{\AA}}
 & \mcc{}
 & \mcc{(ergs cm$^{-2}$ s$^{-1}$)}
 & \mcc{({\AA})}
  \\
\hline
3734.37&H13&-11.865$\pm$0.002&1.839$\pm$0.007\\
3770.63&H11&-12.898$\pm$0.033&0.881$\pm$0.060\\
3819.70&HeI*&-12.579$\pm$0.031&1.280$\pm$0.071\\
3835.38&H9&-12.990$\pm$0.104&0.887$\pm$0.159\\
3868.71&[NeIII]&-11.402$\pm$0.002&2.492$\pm$0.033\\
3889.05&H8&-11.958$\pm$0.002&1.946$\pm$0.036\\
3967.41&[NeIII]&-11.688$\pm$0.003&2.259$\pm$0.037\\
4026.10&HeI*&-12.668$\pm$0.013&1.289$\pm$0.032\\
4101.74&H${\delta}$&-11.796$\pm$0.003&2.145$\pm$0.034\\
4340.47&H${\gamma}$&-11.555$\pm$0.003&2.432$\pm$0.056\\
4363.21&[OIII]&-12.438$\pm$0.023&1.547$\pm$0.079\\
4387.93&HeI&-13.299$\pm$0.150&0.730$\pm$0.185\\
4471.68&HeI&-12.308$\pm$0.031&1.804$\pm$0.165\\
4713.38&HeI*&-12.974$\pm$0.034&1.089$\pm$0.060\\
4740.18&[ArIV]&-13.405$\pm$0.015&0.686$\pm$0.015\\
4861.20&H${\beta}$&-11.163$\pm$0.002&2.885$\pm$0.093\\
4921.93&HeI&-12.845$\pm$0.020&1.201$\pm$0.039\\
4959.52&[OIII]&-10.778$\pm$0.007&3.196$\pm$0.040\\
5007.57&[OIII]&-10.323$\pm$0.001&3.600$\pm$0.043\\
5191.80&[ArIII]&-13.212$\pm$0.100&0.948$\pm$0.152\\
5537.89&[CIIII]&-13.483$\pm$0.007&0.725$\pm$0.003\\
5754.59&[NII]&-13.811$\pm$0.014&0.368$\pm$0.003\\
5875.97&HeI&-11.650$\pm$0.004&2.472$\pm$0.091\\
6300.30&[OI]&-12.688$\pm$0.021&1.511$\pm$0.071\\
6312.10&[SIII]&-13.190$\pm$0.117&0.977$\pm$0.208\\
6363.77&[OI]&-13.103$\pm$0.051&1.119$\pm$0.102\\
6548.05&[NII]&-11.350$\pm$0.004&2.719$\pm$0.160\\
6562.85&H${\alpha}$&-10.590$\pm$0.001&3.489$\pm$0.182\\
6583.45&[NII]&-10.860$\pm$0.002&3.296$\pm$0.082\\
6678.15&HeI&-12.280$\pm$0.005&1.896$\pm$0.057\\
6716.47&[SII]&-13.236$\pm$0.001&1.024$\pm$0.002\\
6730.85&[SII]&-13.116$\pm$0.031&1.094$\pm$0.051\\
7065.71&HeI&-11.831$\pm$0.003&2.354$\pm$0.057\\
7135.80&[ArIII]&-12.056$\pm$0.006&2.128$\pm$0.076\\
 \hline
 \end{tabular}
\end{center}
\end{small}
\end{table*}

\begin{table*}
 \caption{Vy 2-3 fluxes and equivalent widths.}
 \label{tab:19}
\begin{small}
\begin{center}
\begin{tabular}{@{}c@{~}c@{~}c@{~}c@{}}
\hline \hline
   \mcc{Wavelength}
 & \mcc{Ion}
 & \mcc{log [Flux]}
 & \mcc{log [EW]}
  \\
  \mcc{{\AA}}
 & \mcc{}
 & \mcc{(ergs cm$^{-2}$ s$^{-1}$)}
 & \mcc{({\AA})}
  \\
\hline
3750.15&H12&-12.104$\pm$0.010&1.733$\pm$0.025\\
3889.05&H8&-12.559$\pm$0.013&1.317$\pm$0.025\\
4068.91&CIII&-12.630$\pm$0.012&1.270$\pm$0.020\\
4340.47&H${\gamma}$&-12.393$\pm$0.003&1.576$\pm$0.006\\
4363.21&[OIII]&-13.169$\pm$0.020&0.803$\pm$0.025\\
4471.68&HeI&-13.326$\pm$0.080&0.659$\pm$0.094\\
4685.71&HeII&-13.776$\pm$0.073&0.222$\pm$0.079\\
4713.38&HeI*&-13.707$\pm$0.034&0.300$\pm$0.037\\
4861.20&H${\beta}$&-11.997$\pm$0.002&2.047$\pm$0.012\\
4959.52&[OIII]&-11.486$\pm$0.001&2.562$\pm$0.016\\
5007.57&[OIII]&-11.053$\pm$0.000&2.993$\pm$0.010\\
5875.97&HeI&-12.686$\pm$0.016&1.568$\pm$0.043\\
6562.85&H${\alpha}$&-11.436$\pm$0.001&2.938$\pm$0.027\\
6583.45&[NII]&-13.527$\pm$0.009&0.865$\pm$0.017\\
6678.15&HeI&-13.252$\pm$0.011&1.179$\pm$0.019\\
7065.71&HeI&-13.328$\pm$0.015&1.168$\pm$0.025\\
7135.80&[ArIII]&-12.980$\pm$0.003&1.520$\pm$0.008\\
 \hline
 \end{tabular}
\end{center}
\end{small}
\end{table*}
\end{document}